\documentclass[useAMS,usenatbib]{mn2e}

\usepackage{color}
\usepackage{colortbl}
\usepackage{multirow}
\usepackage{dcolumn}
\usepackage{amsmath}
\usepackage{amssymb}
\usepackage{graphicx}
\usepackage{epsfig}
\usepackage{changebar}
\usepackage{natbib}
\usepackage{multirow}
\usepackage{subfigure}
\usepackage{txfonts}
\usepackage{relsize}

\usepackage{longtable,lscape}
\usepackage[figuresright]{rotating}

\newcolumntype{d}[1]{D{.}{\cdot}{#1}}

\newcolumntype{.}{D{.}{.}{-1}}

\newcommand{\lsun}{L$_\odot$}
\newcommand{\msun}{M$_\odot$}

\newcommand{\vlsr}{V$_{\rm{LSR}}$}

\newcommand{\mum}{$\mu$m}

\newcommand{\kms}{km\,s$^{-1}$}

\newcommand{\x}{$\times$}

\newcommand{\poi}{Poisson}

\title[Properties of sites of massive star formation]{The RMS Survey: Distribution and properties of a sample of massive young stars.}
\author[J. S. Urquhart et al.]{
J. S. Urquhart$^{1}$\thanks{E-mail:
James.Urquhart@csiro.au (ATNF)}, T.\,J.\,T.\,Moore$^{2}$, M.\,G.\,Hoare$^{3}$, S.\,L.\,Lumsden$^{3}$,  R.\,D.\,Oudmaijer$^{3}$,
\newauthor J.\,M.\,Rathborne$^{1,4}$, J.\,C.\,Mottram$^{5}$, B.\,Davies$^{3}$, J.\,J.\,Stead$^{3}$\\
$^{1}$Australia Telescope National Facility, CSIRO Astronomy and Space Science, PO Box 76, Epping NSW 1710, Australia \\
$^{2}$Astrophysics Research Institute, Liverpool John Moores University, Twelve Quays House, Egerton Wharf, Birkenhead, CH41\,1LD, UK\\ 
$^{3}$School of Physics and Astrophysics, University of Leeds, Leeds, LS2\,9JT, UK \\
$^{4}$Universidad de Chile, Camino el Observatorio, Las Condes, Santiago, Chile
Casilla 36-D \\
$^{5}$School of Physics, University of Exeter, Exeter, EX7 4QL, UK\\
}

\begin{document}

\date{Accepted ??. Received ??; in original form ??}

\pagerange{\pageref{firstpage}--\pageref{lastpage}} \pubyear{2009}

\maketitle

\label{firstpage}

\begin{abstract}
The Red MSX Source (RMS) survey has identified a large sample of massive young stellar objects (MYSOs) and ultra compact (UC) HII regions from a sample of $\sim$2000 MSX and 2MASS colour selected sources. Using a recent catalogue of molecular clouds derived from the Boston University-Five College Radio Astronomy Observatory (BU-FCRAO) Galactic Ring Survey (GRS), and by applying a Galactic scaleheight cut off of 120\,pc, we solve the distance ambiguity for RMS sources located within  $18\degr < |l| > 54\degr$. These two steps yield kinematic distances to 291 sources out of a possible 326 located within the GRS longitude range. Combining distances and integrated fluxes derived from spectral energy distributions, we estimate luminosities to these sources and find that $>$ 90 per~cent are indicative of the presence of a massive star. We find the completeness limit of our sample is $\sim$10$^4$\,\lsun, which corresponds to a zero age main sequence (ZAMS) star with a  mass of $\sim$12\,\msun. Selecting only these sources, we construct a complete sample of 196 sources.

Comparing the properties of the sample of young massive stars with the general population, we find the RMS-clouds are generally larger, more massive, and more turbulent. We examine the distribution of this sub-sample with respect to the location of the spiral arms and the Galactic bar and find them to be spatially correlated. We identify three significant peaks in the source surface density at Galactocentric radii of approximately 4, 6 and 8\,kpc, which correspond to the proposed positions of the Scutum, Sagittarius and Perseus spiral arms, respectively. Fitting a scale height to the data we obtain an average value of $\sim29\pm0.5$\,pc, which agrees well with other reported values in the literature, however, we note a dependence of the scale height on galactocentric radius with it increases from 30\,pc to 45\,pc between 2.5 and 8.5\,kpc.

\end{abstract}

\section{Introduction}

Throughout their lives massive stars ($M_\star>8$\,\msun) play an important role in many aspects of astrophysics including many of the most energetic phenomena in the Universe. They deposit huge amounts of energy into the interstellar medium in the form of UV radiation, which leads to the creation of HII regions, and kinetic energy via molecular outflows, powerful stellar winds, and supernova explosions. Massive stars are also responsible for the production of enriched material that alters the local chemistry. These feedback processes play an important role in regulating star formation within the surrounding environment, possibly triggering the formation of future generations of stars, and ultimately driving the evolution of their host galaxy (\citealt{kennicutt2005}). 

Given the profound impact massive stars have, not only on their local environment, but also on a Galactic scale, it is crucial to understand the environmental conditions and processes involved in their birth and the earliest stages of their evolution. The life cycle of massive stars once they have emerged from their natal molecular cloud is generally thought to be understood. However, the processes involved in their formation and the early stages of their evolution are still shrouded in mystery, with our understanding lagging behind that of low-mass star formation. 

Massive stars are much rarer than their low-mass counterparts and consequently are generally located much farther away than regions of low-mass star formation.  Furthermore, massive stars evolve extremely rapidly, reaching the main sequence while still deeply embedded within dense cores. The formation of massive stars and the earliest stages of their evolution are therefore hidden behind many magnitudes of visual extinction, only observable at infrared and millimetre wavelengths. Moreover, massive stars are known to form exclusively in clusters (e.g., \citealt{de-wit2004}), which makes it difficult to attribute derived quantities to individual sources. Consequently, the known samples are too small to test many aspects of massive star formation theories, and due to the various selection criteria, are probably unrepresentative of the class as a whole.

We have conducted a new galaxy-wide search (the Red MSX Source, or RMS survey\footnote{http://www.ast.leeds.ac.uk/RMS}) with the specific aim of identifying a large well selected sample of massive young stellar objects (MYSOs).  Observationally, a MYSO is defined as a stellar object which is mid-IR bright and processes sufficient luminosity to form a massive star (greater than a few times 10$^3$\,\lsun) but shows no evidence that a HII region has begun to form (i.e., no radio emission and no spatially extended mid-infrared emission). To place MYSOs in an evolutionary context they are somewhat later than the hot molecular core stage, which is not generally found to be mid-infrared bright (e.g., \citealt{de-buizer2002}), but have not yet begun to ionize their surroundings sufficiently to produce a detectable HII region. Although our efforts are directed towards identifying MYSOs we are also sensitive to the slightly later ultra-compact (UC)\,HII region stage. It is unclear whether core hydrogen burning has begun in the MYSO phase, however, their association with strong molecular outflows indicates accretion is ongoing. As the MYSO evolves and moves towards the main sequence it will begin to form an HII region which will eventually halt accretion at which point the final mass of the star is set. The MYSO and UC\,HII stages are two of the earliest and most important in the evolution of massive stars since they encompass the transition between accretion and UV dominated regimes.

Before we can determine each source's luminosity, and thus, confirm the presence of a massive young star we need to estimate its heliocentric distance. Kinematic distances can be derived using the source velocities in conjunction with a Galactic rotation model (e.g., \citealt{brand1993,alvarez1990,clemens1985}). However, for sources located within the solar circle (Galactocentric radii $<$ 8.5\,kpc) there are two possible kinematic distances for any given velocity, equally spaced on either side of the tangent point, commonly referred to as the near and far distances. Kinematic distance ambiguities affect approximately $\sim$70 per~cent of our sample and therefore represent a serious hurdle that needs to be addressed before many of the important source parameters can be calculated.

In this paper we assign kinematic distances to $\sim$300 young massive star candidates located in the northern Galactic plane by correlating their spatial positions and velocities with a catalogue of molecular clouds, derived from $^{13}$CO data from the Galactic Ring Survey (GRS), for which the near-far distance ambiguities have been solved. In the next section (Sect.\,2) we give a brief overview of both the RMS and GRS surveys and describe the procedure used to identify associations. In Sect.\,\ref{sect:results} we present a summary of the RMS-cloud associations, assign distances, and calculate luminosities for RMS sources. In Sect.\,4 we derive the physical sizes and masses of all GRS clouds and compare the properties of those associated with a sub-sample of the most luminous RMS sources with those that are not currently forming massive stars. We discuss the Galactic distribution of our sample of young massive stars with respect to the positions of the spiral arms and Galactic bar in Sect.\,\ref{sect:discussion}.  In Sect.\,\ref{sect:summary} we present a summary of the results and our conclusions.

\section[]{Survey descriptions and data sets}
 
\subsection{RMS Survey}

The RMS survey is based on a search of the MSX mid-infrared survey of the Galactic Plane (\citealt{price2001}). Using the MSX Point Source Catalogue (\citealt{egan2003}) we colour-selected a large sample of candidate MYSOs and UCHII regions (\citealt{lumsden2002}). This initial sample was further refined using near-infrared photometry obtained from 2MASS (\citealt{cutri2003,skrutskie2006}) to eliminate blue objects, and from visual inspection of the MSX images to remove the more extended sources. Our colour-selection and subsequent filtering produced a sample of $\sim$2000 MYSO candidates (\citealt{lumsden2002}).   

The RMS Survey is a multi-wavelength programme of follow-up observations designed to identify genuine MYSOs and UCHII regions. These observations have been designed to identify and remove other kinds of embedded or dusty objects such as planetary nebulae, evolved stars and nearby low-mass YSOs (\citealt{hoare2005,urquhart2007c}). To date we have used arcsecond resolution mid-infrared imaging from the Spitzer GLIMPSE survey (\citealt{benjamin2003}) or our own ground-based imaging (e.g., \citealt{mottram2007}) to reveal multiple and/or extended sources within the MSX beam, as well as MYSOs in close proximity to existing HII regions. We have obtained arcsecond resolution radio continuum images with the ATCA and the VLA (\citealt{urquhart_radio_south,urquhart_radio_north}) to identify UCHII regions and PNe, whilst observations of  $^{13}$CO transitions (\citealt{urquhart_13co_south, urquhart_13co_north,urquhart2009_h2o}) deliver kinematic velocities. Finally, we use near-infrared spectroscopy (e.g., \citealt{clarke2006}) to distinguish the more pathological evolved stars and confirm source classifications. 

By combining these infrared, millimetre and radio wavelength observations with complementary archival data we have identified approximately 1300 UCHII regions and candidate MYSOs located throughout the Galactic plane ($b<$5\degr; \citealt{urquhart2007c}).

\subsection{The BU-FCRAO Galactic Ring Survey}

The $^{13}$CO ($J$=1--0) Galactic Ring Survey (GRS) covers a total area of 75.4 deg$^2$ between 18\degr $<l<$55.7\degr\ and $|b|<$1\degr\ and is fully sampled with a pixel size of 22\arcsec. The processed data cubes with Galactic longitudes $l<40\degr$ have a \vlsr\ range = $-$5 to 135\,\kms, while those with $l>40\degr$ have \vlsr\ range = $-$5 to 85\,\kms. The spectral resolution  for both velocity ranges is 0.21\,km\,s$^{-1}$.  

In a recent paper, \citet{rathborne2009} used the CLUMPFIND algorithm (\citealt{williams1994}) to analyse this homogeneous, fully sampled data set, resulting in the identification of 829 molecular clouds distributed throughout the GRS region. This work was complemented by  \citet{roman2009} who used archival HI data from the VLA Galactic Plane Survey (VGPS; \citealt{stil2006}) to break the near-far kinematic distance ambiguity towards 750 molecular clouds ($\sim$90 per~cent of all clouds identified). The combination of a complete set of molecular clouds and kinematic distances provides an enormously important catalogue that can be used for determining distances to star-forming regions, allowing their distribution, luminosity functions, etc., to be mapped with respect to the large-scale Galactic structure.

\subsection{Resolving the Kinematic Distance Ambiguity}

Obtaining heliocentric distances to our sources is a crucial element of our campaign as without these we cannot accurately calculate source luminosities, and thus, distinguish between nearby low and intermediate-mass YSOs from the more generally more distance MYSOs. As mentioned in the introduction kinematic distance ambiguities affect a large proportion of our sample ($\sim$900 sources in total) and these need to be resolved before we obtain a robust sample of young massive stars.

There have been a number of studies devoted to resolving distances ambiguities using a variety of methods. \citet{dame1991} found correlations and anti-correlations between near-infrared extinction and molecular clouds traced by CO emission could be used to resolve the ambiguity towards molecular clouds. This method worked by assuming that the anti-correlation was the result of dust associated with foreground clouds. \citet{paladini2004} used a statistical approach to derive a luminosity-physical diameter correlation for HII regions. \citet{downes1980}, \citet{araya2001, araya2002} and \citet{sewilo2004} have used a combination of H110$\alpha$ radio recombination lines (RRLs) and formaldehyde (H$_2$CO) absorption measurements to assign near and far solutions for HII regions. 

Although these studies have had reasonable success in resolving ambiguities they tend to be used only for a particular type of object and/or require a specific set of conditions and so are not universally applicable. Another technique that has been applied successfully to both UCHII regions and YSOs combines 21\,cm HI absorption with a molecular tracer such as $^{13}$CO and implicitly assumes that the source is still embedded within its natal molecular cloud. For HII regions this technique relies on the principle that if a continuum source is located at the near distance, the HI data will not show any absorption from clouds with velocities between that of the HII region and the tangent point, since these are located behind the continuum source with respect to our line of sight. If absorption is present in the HI data at distances between the continuum sources and the velocity of the tangent point the source must be located at the far distance (e.g., \citealt{kuchar1994,kolpak2003}). For YSOs this technique looks for HI self-absorption (SA) at the same velocity as the source and works on the principle that if the source's host cloud is located at the near distance, it would lie in front of a significant column of warmer HI, resulting in absorption by cold HI associated with the cloud. Conversely, the absence of an absorption dip would imply the source is located at the far distance (e.g., \citealt{jackson2002,busfield2006}).

In this study we will focus on the sub-sample of UC\,HII regions and MYSO candidates located within the GRS longitude range. Combining these two data sets has the potential to solve the distance ambiguities towards approximately one-third of \emph{our} inner-Galaxy sources, and 75 per~cent of the sources located in the Northern Galactic Plane. The distance ambiguities in this region have been solved using a combination of the two HI absorption techniques described in the previous paragraph for a flux-limited sample of molecular clouds \citep{roman2009}. Both methods have been used for molecular clouds that are also found to be associated with a HII region with the results of both techniques being check to ensure consistency between the two. The distance solutions have been checked with many of the previous surveys and the results are found to be in reasonable agreement. Therefore the GRS catalogue provides robust distances to a complete sample of molecular clouds located within the first quadrant of the Galactic plane.

\subsection{Matching RMS sources with GRS clouds}

In the Northern Galactic Plane we have 427 RMS sources located within the solar circle that are affected by these kinematic distance ambiguities. Searching this sample we have identified 306 young massive stars (UC\,HIIs and MYSO candidates) located within the GRS region. In order to obtain kinematic velocities for these, we extracted spectra from the GRS data cubes and fitted them with Gaussian profiles. In cases where more than one significant emission peak was present in the CO spectrum, we have used archival maser or high-density gas tracers to determine the source velocity (see \citealt{urquhart_13co_north} for more details). In total, we have been able to assign a unique velocity to 300 MYSO candidates and compact HII regions within the GRS region. 

By comparing the Galactic longitudes and latitudes and the velocities of the RMS sources with those derived for the clouds reported by \citet{rathborne2009} we are able to identify the population of clouds that are responsible for giving birth to the next generation of massive stars in the Galaxy. To find a match we began by searching within the GRS in a small region around the coordinates of each RMS source. This region was 
$5\times 5 \times 11$ resolution elements in the $l \times b \times v$ directions corresponding to 15 arcmin in Galactic longitude and latitude and 4.67 \kms\ in velocity. This  search returned a unique cloud match for 116 
RMS sources. For a further 175 sources, multiple clouds were found within the search radius. In the majority of these case there were only a few pixels from a neighbouring cloud and, to determine a match, we selected the cloud that contributed the most pixels within the search region.

\begin{table}
\begin{center}
\caption{Summary of unmatched RMS sources located within the GRS region.}
\label{tbl:unmatched_sources}
\begin{minipage}{\linewidth}
\begin{tabular}{lrrrr}
\hline
\hline
MSX Name	&\vlsr	& Near Dist.  & Far Dist. & RGC  \\
	&(\kms) 	& (kpc)  & (kpc) &(kpc)  \\

\hline
G019.8922+00.1023	&	45.7	&	3.6	&	12.4	&	5.3	\\
G025.3058+00.5308	&	$\cdots$	&	$\cdots$	&	$\cdots$	&	$\cdots$	\\
G025.4948$-$00.2990	&	$\cdots$	&	$\cdots$	&	$\cdots$	&	$\cdots$	\\
G026.1094$-$00.0944	&	$\cdots$	&	$\cdots$	&	$\cdots$	&	$\cdots$	\\
G027.2220+00.1361$\ast$   &   112.8    	& 6.9			& 8.2			&   3.9        \\
G032.2718$-$00.2260	&	22.3	&	1.6	&	12.8	&	7.2	\\
G034.6243$-$00.1300	&	23.0	&	1.6	&	12.4	&	7.2	\\
G039.4943$-$00.9933$\dagger$	&	53.2	&	3.6	&	9.5	&	6.2	\\
G039.8651+00.6474	&	$\cdots$	&	$\cdots$	&	$\cdots$	&	$\cdots$	\\
G042.0977+00.3521	&	20.7	&	1.4	&	11.2	&	7.5	\\
G043.9674+00.9938$\ddagger$	&	$-$19.7	&	14.1	&	14.1	&	10.0	\\
G047.0998+00.4799	&	$\cdots$	&	$\cdots$	&	$\cdots$	&	$\cdots$	\\
G047.9002+00.0671$\ast$	&	72.0	&	5.3	&	5.3	&	5.9	\\
G052.7528+00.3343	&	15.2	&	1.0	&	9.2	&	7.9	\\
G053.9584+00.0317	&	41.6	&	4.1	&	5.9	&	6.9	\\

\hline\\
\end{tabular}\\
$\dagger$ Source assigned to the near distance as the far distance would imply a source location more than 4 times the Galactic scale height from the mid-plane and is thus unlikely.

$\ddagger$ Source velocity lies outside the velocity range covered by the GRS.

$\ast$ This source is located within $\sim$0.5\,kpc of the tangent point and has been assigned the tangent point velocity.

\end{minipage}
\end{center}
\end{table}

We failed to find any cloud association for fifteen  of the 306 RMS sources; we present a list of these sources in 
Table\,\ref{tbl:unmatched_sources}. Of these, the GRS molecular line observations towards two failed to detected any significant level of emission in the observed velocity range. The CO results towards four sources are inconclusive  with multiple components with similar line strengths being detected and no other tracer is available. We note that the velocity assigned to G043.9674+00.9938 was derived from the CS observations presented by \citet{bronfman1996} and falls outside the velocity range covered by the GRS, which explains why we were unable to find a match for this particular source. The limited velocity range covered by the GRS survey may also explain the lack of any detected CO emission towards the two sources towards which no CO emission is detected. 

This leaves eight sources with assigned velocities but for which no associated cloud has been identified. One possible explanation for this is that an incorrect velocity has been assigned to the RMS source. Another possibility is that the host cloud has not been detected by the CLUMPFIND algorithm. To investigate these possibilities we compared the $^{13}$CO spectra with other molecular transitions reported in the literature towards these sources to check the velocity assignment was consistent with previous studies. For seven  sources we find  CS ($J$=2--1) emission (\citealt{bronfman1996}) at the same velocity as the $^{13}$CO emission. The CS rotational transition has a much higher critical density than the $^{13}$CO transition, and thus,  the presence of the two transitions at the same velocity makes it unlikely the assigned velocities are incorrect. 

\begin{figure}
\includegraphics[width=0.49\textwidth, trim= 50 0 0 0]{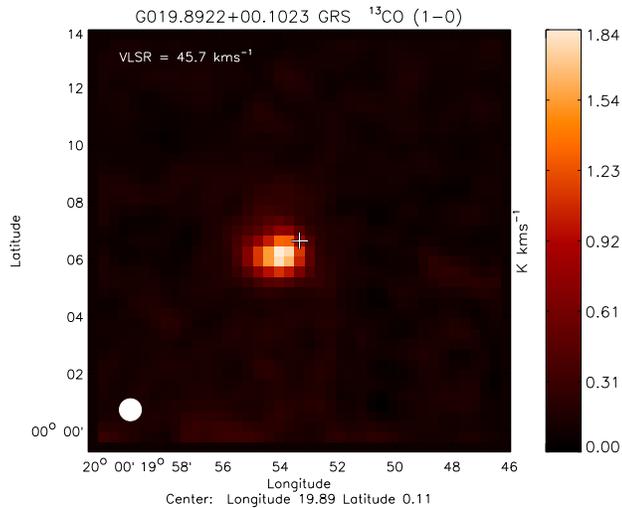}

\caption{\label{fig:integrated_map} Integrated $^{13}$CO $J$=1--0 emission in a 15\arcmin \x 15\arcmin\ region around G019.8922+00.1023. The position of the RMS source is indicated by a cross. The image was produced by integrating the GRS cube over a velocity range of three times the velocity dispersion either side of the source's assigned velocity. The FCRAO beam size is shown in the bottom left hand corner.} 
\end{figure}

Having confirmed that the velocities assigned to the RMS sources are likely to be correct we need to investigate why we have been unable to associate them with a molecular cloud. In order to investigate the spatial extent of the emission surrounding each of these sources we produced integrated velocity maps of the $^{13}$CO emission around each source at the assigned velocity. In Fig.\,\ref{fig:integrated_map} we present an example of one of these integrated $^{13}$CO maps showing the emission coincident with RMS source G019.8922+00.1023 in both position and velocity. This reveals a relatively bright, compact region of emission coincident with the the RMS source. We find a similar situation for the other sources, however, the $^{13}$CO emission in nearly all cases is quite weak ($T_{\rm{A}}^* \sim 0.4-1.2$\,K) and it is therefore likely that the emission associated with these small clouds fell below the criteria used to generate the GRS catalogue (\citealt{rathborne2009}).

\begin{figure}
\includegraphics[width=0.45\textwidth]{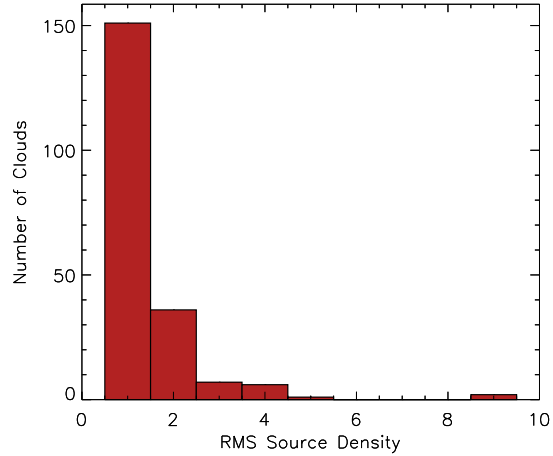}
\caption{\label{fig:sources_per_cloud} The number of RMS sources found in each molecular cloud.}
\end{figure}

\section[]{Distances and luminosities}
\label{sect:results}

\citet{roman2009} used the \citet{clemens1985} Galactic rotation model to determine the kinematic distances to the GRS sample of clouds. However, to maintain consistency with previous published work on the RMS sample (i.e., \citealt{urquhart_13co_south,urquhart_13co_north}) we have recalculated their distances using the \citet{brand1993} rotation model assuming a distance to the Galactic centre of 8.5\,kpc and a solar velocity of 220\,\kms. The difference in derived distances between the two models is small (typically $\pm$0.3\,kpc) compared to the error associated with kinematic distances, which are of order $\pm$1\,kpc in this part of the Galaxy, and therefore the actual choice of rotation curve does not have any impact our findings.

\subsection{RMS-GRS cloud associations}

We have identified the molecular clouds associated with 291 RMS sources. In Table\,\ref{tbl:cloud_summary} we present a summary of the GRS cloud parameters and the number of associated young massive stars  (only a small portion of the table is provided here, the full table is only available in electronic form at the CDS via anonymous ftp to cdsarc.u-strasbg.fr (130.79.125.5) or via http://cdsweb.u-strasbg.fr/cgi-bin/qcat?J/A+A/). Given the angular size of the molecular clouds identified by \citet{rathborne2009} we might expect to find multiple regions of star formation within a significant number of clouds and this is indeed the case. We find the RMS sources are associated with 204 clouds, with 53 clouds being found to be associated with two or more sites of star formation. 

In Fig.\,\ref{fig:sources_per_cloud} we present a histogram of the number of RMS sources found in each cloud. The majority of clouds are associated with a single RMS source, with the number of clouds falling off rapidly as a function of RMS source density. We find two clouds that have an RMS source count of nine (G043.19$-$00.01 and G049.49$-$00.41; \citealt{rathborne2009}). Being associated with so many potential sites of massive star formation distinguishes these clouds as two of the most intense regions of massive star formation in this part of the Galaxy. G049.49$-$00.41  is part of the  W49 giant molecular cloud, which is the most massive cloud in the Galaxy outside the Galactic centre (\citealt{simon2001}). The second cloud, G043.19$-$00.01, is coincident with a well known OH maser source (OH\,43.8$-$0.1; \citealt{braz1983}) and a variable H$_2$O maser (\citealt{lekht2000}). 

\citet{roman2009} resolved the distance ambiguities to a hundred and eighty-five of the RMS-GRS cloud associations, which are themselves host to a total of 268 RMS sources. We are therefore able to assign kinematic distances to $\sim$90 per~cent of the matched RMS sources. Furthermore, examination of the remaining RMS sources in the GRS region without an assigned cloud identified four sources for which a distance can be assigned (see superscripts and footnotes in Table\,\ref{tbl:unmatched_sources} for individual source details). One of these (G043.9674+00.9938) has a velocity outside the GRS range that corresponds to a location outside the solar circle, and therefore is not subject to a kinematic distance ambiguity. Another source (G025.3981+00.5617) is assigned to the near distance through an analysis of its heights above the Galactic mid-plane (see following sub-section for details). Finally, we find two sources that are located within $\sim$0.5\,kpc of the tangent point and so have been assigned the distance of the tangent point. Including these four sources, we have resolved the distance ambiguities towards 272 RMS sources.

\begin{table}

\begin{center}
\caption{\label{tbl:cloud_summary} Summary of GRS cloud parameters and RMS source associations.}
\begin{minipage}{\linewidth}
\begin{tabular}{l.....}
\hline
\hline

GRS	Cloud	&	\multicolumn{1}{c}{Radii} 				& \multicolumn{1}{c}{$T_{\rm{ex}}$} 			&\multicolumn{1}{c}{$\tau$} &	\multicolumn{1}{c}{$M_{\rm{LTE}}$}		&	\multicolumn{1}{c}{No. RMS}  \\
Name				&	\multicolumn{1}{c}{(pc)} 		& \multicolumn{1}{c}{(K)} 		& 					&\multicolumn{1}{c}{Log$_{10}$($M$ (\msun))}     &\multicolumn{1}{c}{Sources}   \\
\hline
G021.34+00.84	&	\multicolumn{1}{c}{$\cdots$}	&	7.01	&	0.14	&	\multicolumn{1}{c}{$\cdots$}	&	\multicolumn{1}{c}{$\cdots$}	\\
G021.34$-$00.16	&	20.3	&	9.12	&	0.10	&	4.13	&	\multicolumn{1}{c}{$\cdots$}	\\
G021.39$-$00.01	&	21.5	&	9.49	&	0.12	&	4.35	&	\multicolumn{1}{c}{$\cdots$}	\\
G021.39$-$00.26	&	29.0	&	7.68	&	0.13	&	4.46	&	2	\\
G021.39$-$00.56	&	18.9	&	10.10	&	0.09	&	4.08	&	\multicolumn{1}{c}{$\cdots$}	\\
G021.44+00.44	&	\multicolumn{1}{c}{$\cdots$}	&	7.53	&	0.11	&	\multicolumn{1}{c}{$\cdots$}	&	\multicolumn{1}{c}{$\cdots$}	\\
G021.44$-$00.71	&	13.8	&	9.99	&	0.12	&	3.82	&	\multicolumn{1}{c}{$\cdots$}	\\
G021.49$-$00.26	&	15.9	&	7.48	&	0.14	&	3.17	&	\multicolumn{1}{c}{$\cdots$}	\\
G021.49$-$00.71	&	45.0	&	8.25	&	0.10	&	4.60	&	\multicolumn{1}{c}{$\cdots$}	\\
G021.54+00.24	&	36.1	&	7.64	&	0.12	&	4.40	&	\multicolumn{1}{c}{$\cdots$}	\\
G021.54$-$00.01	&	37.7	&	9.91	&	0.10	&	4.59	&	\multicolumn{1}{c}{$\cdots$}	\\
G021.54$-$00.11	&	18.8	&	9.27	&	0.11	&	3.70	&	\multicolumn{1}{c}{$\cdots$}	\\
G021.59+00.49	&	4.7	&	12.05	&	0.08	&	2.73	&	\multicolumn{1}{c}{$\cdots$}	\\
G021.59$-$00.06	&	15.4	&	7.68	&	0.12	&	3.45	&	\multicolumn{1}{c}{$\cdots$}	\\
G021.59$-$00.06	&	26.8	&	7.68	&	0.12	&	4.31	&	1	\\
G021.59$-$00.11	&	26.9	&	7.31	&	0.15	&	4.09	&	\multicolumn{1}{c}{$\cdots$}	\\
G021.64+00.14	&	2.9	&	6.96	&	0.16	&	1.67	&	\multicolumn{1}{c}{$\cdots$}	\\
G021.69+00.09	&	14.6	&	8.55	&	0.09	&	3.37	&	\multicolumn{1}{c}{$\cdots$}	\\
G021.69+00.59	&	\multicolumn{1}{c}{$\cdots$}	&	8.50	&	0.10	&	\multicolumn{1}{c}{$\cdots$}	&	\multicolumn{1}{c}{$\cdots$}	\\
G021.69$-$00.01	&	19.0	&	7.12	&	0.14	&	3.92	&	1	\\
G021.69$-$00.06	&	43.6	&	8.68	&	0.10	&	4.55	&	\multicolumn{1}{c}{$\cdots$}	\\
G021.74+00.09	&	53.7	&	8.47	&	0.10	&	4.63	&	\multicolumn{1}{c}{$\cdots$}	\\
G021.74$-$00.01	&	24.4	&	9.97	&	0.09	&	4.18	&	\multicolumn{1}{c}{$\cdots$}	\\
G021.84+00.59	&	18.2	&	5.96	&	0.21	&	3.23	&	\multicolumn{1}{c}{$\cdots$}	\\
G021.84$-$00.26	&	23.1	&	10.54	&	0.09	&	4.44	&	\multicolumn{1}{c}{$\cdots$}	\\
G021.89$-$00.56	&	30.2	&	9.37	&	0.11	&	4.92	&	\multicolumn{1}{c}{$\cdots$}	\\
G021.94+00.04	&	13.0	&	6.60	&	0.16	&	3.56	&	\multicolumn{1}{c}{$\cdots$}	\\
G021.94+00.14	&	34.8	&	8.76	&	0.11	&	4.73	&	\multicolumn{1}{c}{$\cdots$}	\\
G021.99$-$00.06	&	32.6	&	7.72	&	0.11	&	4.18	&	\multicolumn{1}{c}{$\cdots$}	\\
G022.04+00.19	&	26.2	&	10.35	&	0.10	&	4.54	&	\multicolumn{1}{c}{$\cdots$}	\\
\hline
\end{tabular}\\
Notes: Only a small portion of the data is provided here, the full table is only  available in electronic form at the CDS via anonymous ftp to cdsarc.u-strasbg.fr (130.79.125.5) or via http://cdsweb.u-strasbg.fr/cgi-bin/qcat?J/A+A/.
\end{minipage}
\end{center}

\end{table}

\subsection{Distances to high latitude sources}

\begin{figure}
\includegraphics[width=0.45\textwidth]{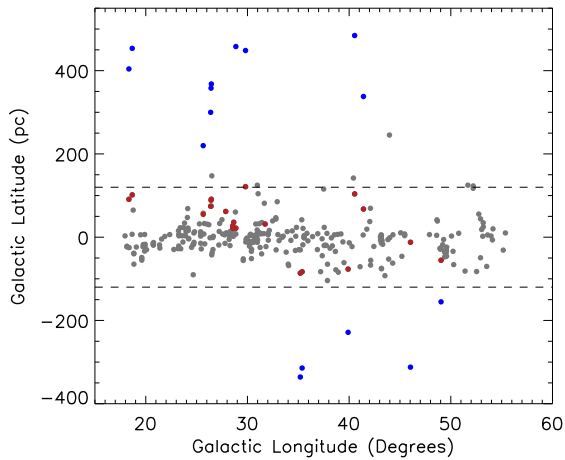}
\caption{\label{fig:kda_plot} Distribution of heights above the mid-plane as a function of Galactic longitude for RMS sources with assigned distances (grey) and the heights that correspond to the near and far distances (shown in red and blue respectively) for high latitude sources (i.e., $|b| > 1\degr$).}
\end{figure}

In the previous section we matched RMS sources with molecular clouds identified within the GRS and obtained distances to 272 young massive stars. However, in order to obtain a fully representative sample we also need to resolve the distance ambiguities to sources located within the GRS longitude range but with $|b| > 1$\degr. Although we have obtained $^{13}$CO velocities towards a further twenty sources in this category, there are currently no archival HI data available at their Galactic latitudes, and thus, we cannot employ either the HI self-absorption or continuum absorption methods used by \citet{roman2009}. However, the larger angular separation of these sources from the Galactic mid-plane would correspond to unrealistically large distances from the plane if a source were located at the far distance. 

In Fig.\,\ref{fig:kda_plot} we plot the height above the Galactic mid-plane for the 20 sources with $|\,b\,| >1$\degr, corresponding to the near and far distances. In addition to the high latitude sources, we also show the distribution of 
heights of sources with assigned distances, and a scale that corresponds to 4 times the Galactic scale-height 
(see the discussion in Sect.\,\ref{sect:discussion}). Comparing the heights of the RMS sources with assigned distances with those of the high latitude sources, it is very unlikely that any of them are located at the far distance. We have therefore assigned the near distance to these sources.

Using a combination of association with molecular clouds of known distance, and applying the scale-height cutoff described in the previous paragraph, we have been able to assign distances to 292 RMS sources from our initial sample of 326 located within the GRS longitude range. This corresponds to $\sim$90 per~cent of the sample.

 \subsection{Luminosity}

Luminosities have been calculated for each source with an assigned distance by fitting a spectral energy distribution (SED) to a set of infrared to millimetre flux measurements. Near- and mid-infrared fluxes have been taken from the 2MASS and MSX point source catalogues. Far-infrared fluxes were obtained from either MIPSGAL or  the IRAS Galaxy Atlas (IGA; \citealt{cao1997}) images, using a set of 2-D background-fitting aperture photometry routines (see \citealt{mottram2009} for details). The MIPSGAL data are generally superior to the IRAS data although, in many cases, the MIPSGAL images were found to be saturated at the location of the RMS sources and could not be used. Finally, (sub)millimetre fluxes were obtained from archival SCUBA (\citealt{di-francesco2008}), SIMBA (\citealt{faundez2004,hill2005,beltran2006}) and MAMBO (\citealt{beuther2002}) data.

\begin{figure}
\includegraphics[width=0.45\textwidth]{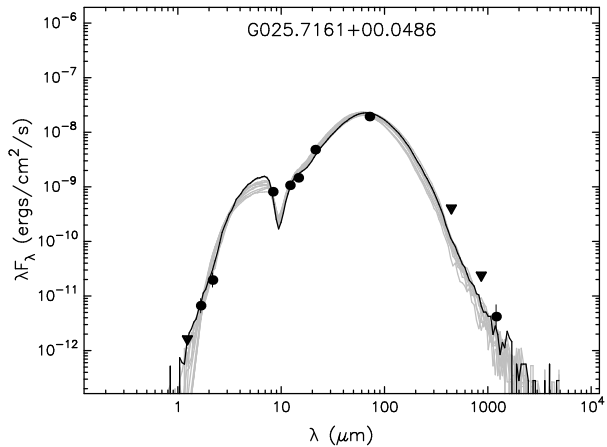}
\caption{\label{fig:sed_fit} An example of a SED fit to infrared and submillimetre fluxes. The flux values and their associated errors are shown as circles and vertical lines respectively, upper limits are shown as inverted triangles. The solid black line shows the best fit to the data while the grey lines trace the next ten best SED fits to the data.}
\end{figure}

SEDs were constructed for each source using the model fitter produced by \citet{robitaille2006,robitaille2007}. In Fig.\,\ref{fig:sed_fit} we present the measured fluxes and subsequent SED fit to the data obtained for the RMS source G025.7161+00.0486. Source luminosities were calculated by integrating under the SED and multiplying by the square of the assigned distance.  A detailed description of the photometry routines used to extract fluxes from the MIPSGAL and IRAS images, and a full discussion of the procedures will be presented in an upcoming paper (Mottram et al. 2010, submitted).

\begin{figure*}
\begin{center}
\includegraphics[width=0.90\textwidth]{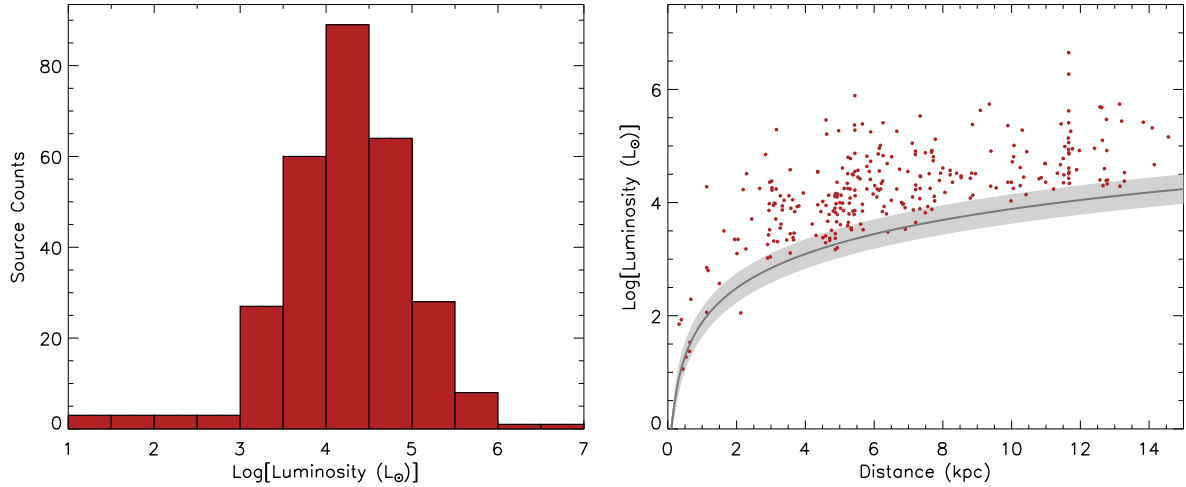}
\end{center}
\caption{\label{fig:luminosities}  Left panel: Distribution of RMS source luminosities. The bin size used is 0.5 dex. Right panel: Luminosity distribution as a function of heliocentric distance. The dark line and light grey shaded region indicates the limiting sensitivity of the MSX 21\,\mum\ band and its associated uncertainty.}

\end{figure*}

In Fig.\,\ref{fig:luminosities} we present the luminosity distribution of all the RMS sources located within the GRS longitude range that have assigned distances. The left panel contains a histogram of the source luminosity function which peaks at approximately $10^4$\,\lsun, confirming that we are finding a significant number of young massive stars. The latest stellar spectral type that can ionize its surroundings and form an HII region is a B3 which has a corresponding mass of $\sim$5.2\,\msun\ and a luminosity of $\sim$550\,\lsun\ (\citealt{martins2005, meynet2000}), and thus, we define sources with luminosities larger than this as a massive star. We find that 290  of the 292 RMS sources ($\sim$99 per~cent) with assigned distances have larger luminosities than would be expected from a B3 star --- assuming the observed luminosity arises from a single embedded star. Massive stars are known to form exclusively in clusters. However, \citet{wood1989} showed that for a realistic initial mass function the most massive member of a cluster is only a couple of spectral classes lower than that determined by assuming the luminosity was due to a single star. Taking this into account we find that any RMS source with a luminosity greater than 10$^4$\,\lsun\ (i.e., equivalent to a star of spectral type B1 or earlier) is likely to contain a massive star. Applying this limit to identify a bona fide sample of massive young stars we find approximately two-thirds (193) of our sources have luminosities over 10$^4$\,\lsun, and are therefore luminous enough to host massive young stars. We note that these fractions only apply to the current sample since there is likely to be number of high mass stars that fall below our detection threshold. 

In the right panel of Fig.\,\ref{fig:luminosities} we present a plot showing the distribution of source luminosities as a function of heliocentric distance along with the MSX 21\,\mum\ limiting sensitivity. The sensitivity has been estimated by first calculating the flux in the MSX 21\,\mum\ (band E) using $F_E=4.041\times10^{-14}\,S_{21}$\,Wm$^2$ (\citealt{cohen2000}), where $S_{21}$ is the MSX 21\,\mum\ detection limit ($\sim$2.7\,Jy). This value is then multiplied by a factor of 24, calculated from the ratio of bolometric fluxes, determined from SED fits, to the MSX band E fluxes (Mottram et al. 2010, submitted). The fairly homogeneous distribution in heliocentric distance of RMS sources above 10$^4$\,\lsun, would suggest that our sample of young massive stars is complete for luminosities above this across the Galaxy. To properly determine the completeness limit requires modelling of the luminosity function which is beyond the scope of this paper and will be the subject of a subsequent paper (Davies et al. 2010, in prep.), however, preliminary results are consistent with the completeness level derived from  Fig.\,\ref{fig:luminosities} . In Table\,\ref{tbl:rms_summary} we present a summary of each RMS source located within the GRS longitude range including positions, associated GRS cloud name, assigned kinematic distance, height above the mid-plane, Galactocentric distance and bolometric luminosity (only a small portion of the table is provided here, the full table is only available in electronic form at the CDS via anonymous ftp to cdsarc.u-strasbg.fr (130.79.125.5) or via http://cdsweb.u-strasbg.fr/cgi-bin/qcat?J/A+A/). 

\begin{table*}

\begin{center}
\caption{\label{tbl:rms_summary} Summary of RMS source parameters located within the GRS longitude range.}

\begin{minipage}{\linewidth}
\begin{tabular}{lccl....}
\hline
\hline

MSX Name		&	RA 				& Dec. 			&GRS Cloud	Name		&\multicolumn{1}{c}{Distance}  & \multicolumn{1}{c}{Height} &	\multicolumn{1}{c}{R$_{\rm{GC}}$} & \multicolumn{1}{c}{Luminosity} \\
				&	(J2000) 		& (J2000) 		& 					&\multicolumn{1}{c}{(kpc)}     & \multicolumn{1}{c}{(pc)}   &	\multicolumn{1}{c}{(kpc)} & \multicolumn{1}{c}{Log$_{10}$(\lsun)} \\
\hline
G017.9642+00.0798	&	18:23:20.82	&	$-$13:15:05.4	&	G017.99+00.09	&	2.27	&	3.2	&	6.4	&	3.2	\\
G017.9789+00.2335	&	18:22:49.08	&	$-$13:09:59.0	&	G017.99+00.09	&	1.50	&	6.1	&	7.1	&	2.6	\\
G018.1409$-$00.3021	&	18:25:04.46	&	$-$13:16:26.7	&	G018.24$-$00.41	&	4.36	&	-23.0	&	4.6	&	4.5	\\
G018.1504$-$00.2794	&	18:25:00.62	&	$-$13:15:18.3	&	G018.24$-$00.41	&	4.36	&	-21.2	&	4.6	&	4.5	\\
G018.3029$-$00.3910	&	18:25:42.48	&	$-$13:10:20.2	&	G018.34$-$00.36	&	2.98	&	-20.4	&	5.7	&	4.7	\\
G018.3412+01.7681	&	18:17:58.15	&	$-$12:07:24.9	&	\multicolumn{1}{c}{$\cdots$}	&	2.96	&	91.3	&	5.8	&	4.4	\\
G018.3706$-$00.3818	&	18:25:48.26	&	$-$13:06:29.5	&	G018.39$-$00.41	&	3.67	&	-24.5	&	5.1	&	3.5	\\
G018.6608+00.0372	&	18:24:50.42	&	$-$12:39:20.8	&	G018.69+00.04	&	10.86	&	7.0	&	3.9	&	4.4	\\
G018.6696+01.9631	&	18:17:54.09	&	$-$11:44:32.2	&	\multicolumn{1}{c}{$\cdots$}	&	2.95	&	101.1	&	5.8	&	4.0	\\
G018.7621+00.2634	&	18:24:12.91	&	$-$12:27:37.8	&	G018.64+00.29	&	14.09	&	64.8	&	6.6	&	5.3	\\
G018.8246$-$00.4673	&	18:26:59.01	&	$-$12:44:47.0	&	G018.94$-$00.26	&	4.60	&	-37.5	&	4.4	&	3.9	\\
G018.8319$-$00.4788	&	18:27:02.35	&	$-$12:44:43.0	&	G018.89$-$00.51	&	4.65	&	-38.9	&	4.4	&	3.7	\\
G018.8330$-$00.3004	&	18:26:23.64	&	$-$12:39:40.3	&	G018.69$-$00.06	&	12.41	&	-65.1	&	5.2	&	5.0	\\
G019.0741$-$00.2861	&	18:26:48.14	&	$-$12:26:28.3	&	G019.09$-$00.26	&	4.62	&	-23.1	&	4.4	&	5.2	\\
G019.4902+00.1350	&	18:26:04.26	&	$-$11:52:36.1	&	G019.39$-$00.01	&	2.44	&	5.8	&	6.2	&	3.7	\\
G019.6053$-$00.9012	&	18:30:02.59	&	$-$12:15:23.7	&	G019.59$-$00.91	&	3.07	&	-48.2	&	5.7	&	3.9	\\
G019.6085$-$00.2357	&	18:27:38.30	&	$-$11:56:41.2	&	G019.64$-$00.26	&	12.57	&	-51.7	&	5.4	&	5.7	\\
G019.6106$-$00.2531	&	18:27:42.28	&	$-$11:57:03.9	&	G019.64$-$00.26	&	12.57	&	-55.5	&	5.4	&	4.4	\\
G019.7268$-$00.1132	&	18:27:25.20	&	$-$11:46:59.1	&	G019.59$-$00.06	&	11.67	&	-23.1	&	4.7	&	4.5	\\
G019.7403+00.2799	&	18:26:01.48	&	$-$11:35:16.4	&	G019.24+00.34	&	1.90	&	9.3	&	6.7	&	3.2	\\
G019.7540$-$00.1279	&	18:27:31.48	&	$-$11:45:56.8	&	G019.59+00.09	&	8.00	&	-17.9	&	2.9	&	4.6	\\
G019.8817$-$00.5347	&	18:29:14.35	&	$-$11:50:30.1	&	G019.89$-$00.56	&	3.52	&	-32.8	&	5.3	&	4.1	\\
G019.8922+00.1023	&	18:26:57.28	&	$-$11:32:10.6	&	\multicolumn{1}{c}{$\cdots$}	&	\multicolumn{1}{c}{$\cdots$}	&	\multicolumn{1}{c}{$\cdots$}	&	\multicolumn{1}{c}{$\cdots$}	&	\multicolumn{1}{c}{$\cdots$}	\\
G019.9224$-$00.2577	&	18:28:18.83	&	$-$11:40:37.5	&	G019.94$-$00.21	&	4.59	&	-20.6	&	4.5	&	3.3	\\
G019.9386$-$00.2079	&	18:28:09.88	&	$-$11:38:22.9	&	G019.94$-$00.21	&	4.59	&	-16.6	&	4.5	&	3.4	\\
G020.0801$-$00.1360	&	18:28:10.41	&	$-$11:28:51.6	&	G019.74$-$00.26	&	12.62	&	-30.0	&	5.5	&	5.7	\\
G020.5143+00.4936	&	18:26:43.51	&	$-$10:48:12.6	&	G020.49+00.49	&	\multicolumn{1}{c}{$\cdots$}	&	\multicolumn{1}{c}{$\cdots$}	&	\multicolumn{1}{c}{$\cdots$}	&	\multicolumn{1}{c}{$\cdots$}	\\
G020.7121$-$00.0754	&	18:29:09.09	&	$-$10:53:35.1	&	G020.79$-$00.06	&	11.66	&	-15.3	&	4.8	&	4.4	\\
G020.7438$-$00.0952	&	18:29:16.96	&	$-$10:52:27.1	&	G020.79$-$00.06	&	11.66	&	-19.4	&	4.8	&	4.9	\\
G020.7491$-$00.0898	&	18:29:16.39	&	$-$10:52:01.2	&	G020.79$-$00.06	&	11.66	&	-18.3	&	4.8	&	5.1	\\
\hline
\end{tabular}\\
Notes: Only a small portion of the data is provided here, the full table is only  available in electronic form at the CDS via anonymous ftp to cdsarc.u-strasbg.fr (130.79.125.5) or via http://cdsweb.u-strasbg.fr/cgi-bin/qcat?J/A+A/.
\end{minipage}
\end{center}

\end{table*}

\begin{figure*}
\begin{center}
\includegraphics[width=0.90\textwidth]{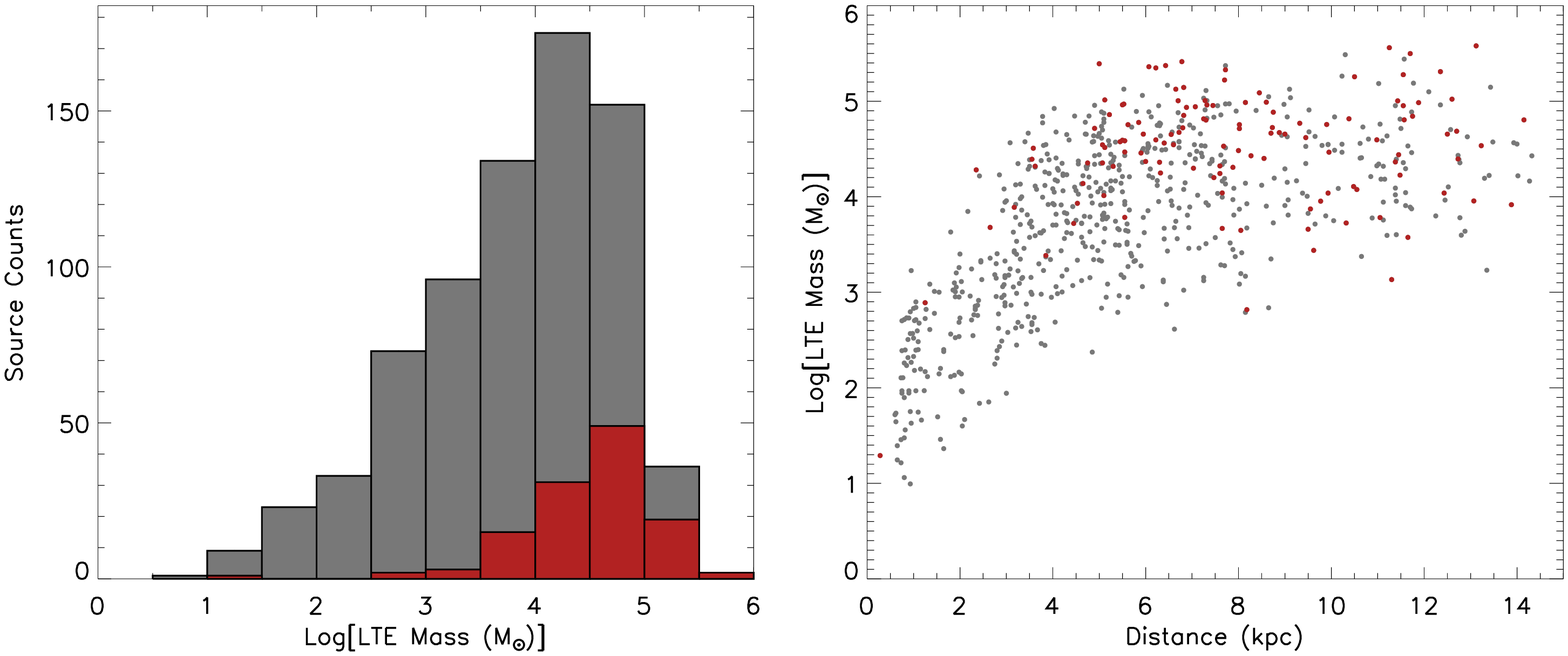}
\end{center}
\caption{\label{fig:grs_masses}  Left panel: Distribution of GRS cloud LTE masses. The bin size used is 0.5 dex. Right panel: LTE mass distribution as a function of heliocentric distance. In both plots the whole sample of GRS clouds is shown in grey while the GRS clouds associated with an RMS source with a luminosity greater than 10$^4$\,\lsun\ are shown in red.}

\end{figure*}

\section{Cloud properties}

In this section we will discuss the Galactic distribution of molecular gas and compare the physical properties of clouds found to be associated with the RMS sample of young massive stars. To facilitate these discussions we need to calculate a few key parameters currently unavailable in either of the \citet{roman2009} or \citet{rathborne2009} papers. 

\subsection{Deriving cloud masses and physical sizes}
\label{sect:masses}

In this subsection we will estimate masses and physical radii for every cloud that has an unambiguous distance, and for which an excitation temperature has been calculated by \citet{rathborne2009}. The observed radius of each cloud has been estimated using the projected area of the cloud on the sky, $A$, given by \citet{rathborne2009} via:

\begin{eqnarray}
R_{\rm{obs}}=\left(\frac{A}{\pi}\right)^{0.5}
\end{eqnarray}

\noindent To correct for the spatial resolution we use the following:

\begin{eqnarray}
R_{\rm{clump}}=\left[R_{\rm{obs}}^2-\left(\frac{\theta_{\rm{FWHM}}} {2}\right)^2\right]^{0.5}
\end{eqnarray}

\noindent where $\theta_{\rm{FWHM}}$ is the effective beam size of the smoothed GRS cube.

\citet{rathborne2009} calculated the optical depths and excitation temperatures for all of the clouds by combining $^{12}$CO ($J$=1--0) data from the University of Massachusetts-Stony Brook survey (\citealt{sanders1986}) and the $^{13}$CO data from the GRS. They calculated typical optical depths and excitation temperatures of $\sim$0.13 and 9\,K respectively. We can now use these results to estimate the mass of the individual clouds. Following the method outlined by \citet{bourke1997} it is possible to estimate the total mass of a cloud using a single optically thin molecular transition and assuming an appropriate abundance ratio. Since only a single transition is used, the mass is determined assuming local thermodynamic equilibrium (LTE) and 
is referred to hereafter as the LTE mass ($M_{\rm{LTE}}$). We use the following equation taken from \citet{bourke1997}:

\begin{eqnarray}
\frac{M_{{\rm{LTE}}}}{M_\odot}&=&\frac{\mu_{\rm{m}}} {2.72m_{\rm{H}}} \frac{[{\rm{H_2}}/^{13}{\rm{CO}}]}{7 \times 10^5} \left(\frac{D}{{\rm{kpc}}}\right)^2 \nonumber \\
&&\times \frac{0.312}{1-{\rm{exp}}(-5.29/T_{\rm{ex}})} \int \int T_{\rm{mb}}\ {\rm{d}}v\    {\rm{d}}\Omega 
\end{eqnarray}

\noindent where the $\mu_{\rm{m}}$ is the mean molecular mass per H$_2$ molecule,  $m_{\rm{H}}$ is atomic mass of hydrogen, $D$ is the heliocentric distance, $v$ is in \kms\ and $\Omega$ is the source solid angle in arcmin$^2$. We have assumed an abundance ratio of H$_2$ to $^{13}$CO of $7 \times 10^5$ (\citealt{frerking1982,bourke1997}) a mean molecular mass of 1.36 $m_{\rm{H}}$, taking account of the H$_{\rm{e}}$ content of of the gas. Distances were taken from \citet{roman2009} and the $T_{\rm{ex}}$ from \citet{rathborne2009}. The $\int \int T_{\rm{mb}}\ {\rm{d}}v\    {\rm{d}}\Omega$ term in Eqn.\,3 is equal to the total integrated intensity which has been calculated for each cloud and presented in \citet{rathborne2009}. We present the derived values for cloud LTE masses in Col.\,5 of Table \ref{tbl:cloud_summary}.

We have estimated the LTE masses for 734 GRS clouds. In the left and right panels of Fig.\,\ref{fig:grs_masses} we present the distribution of LTE cloud mass and plot the distribution of cloud mass as a function of heliocentric distance. In both of these plots we show the whole catalogue of clouds in grey and indicate the clouds associated with RMS sources with a luminosity above 10$^4$\,\lsun\ in red. 

The mass distribution shown in the left panel of Fig.\,\ref{fig:grs_masses} peaks at masses $\sim$10$^{4-4.5}$\,\msun, which is smaller than the masses generally reported for giant molecular clouds (GMCs; $\sim$10$^{5-6}$\,\msun; \citealt{solomon1987}). The lower masses derived for the GRS clouds are consequence of using the weaker $^{13}$CO transition rather than the $^{12}$CO transition normally used to determine the properties of GMCs. We are therefore not tracing the full extent of the GMCs and missing a significant amount of emission from the more extended low-density envelope. We can obtain an estimate of the proportion of the GMCs traced by the $^{13}$CO emission by making a few simple assumptions. In the lower panel of Fig.\,7 we present a histogram of the mass density per kpc$^{-2}$ as a function of galactocentric radius. Assuming that the region sampled by the GRS is representative of the Galaxy as a whole and integrating the emission over 2$\pi$ we estimate the total mass of molecular gas would be $6.3 \times 10^7$\,\msun. Comparing the mass estimated from the $^{13}$CO emission with the value of $\sim 9 \times 10^8$\,\msun\ derived by \citet{blitz1999} for the total mass in GMCs (based on work by \citet{dame1993} and \citet{hunter1997} reveals that only a small fraction ($\sim$7~per~cent) of the entire GMC has sufficient column density to have been detected by the GRS. 

\subsection{Galactic distribution of molecular gas}

\begin{figure}

\includegraphics[width=0.9\linewidth]{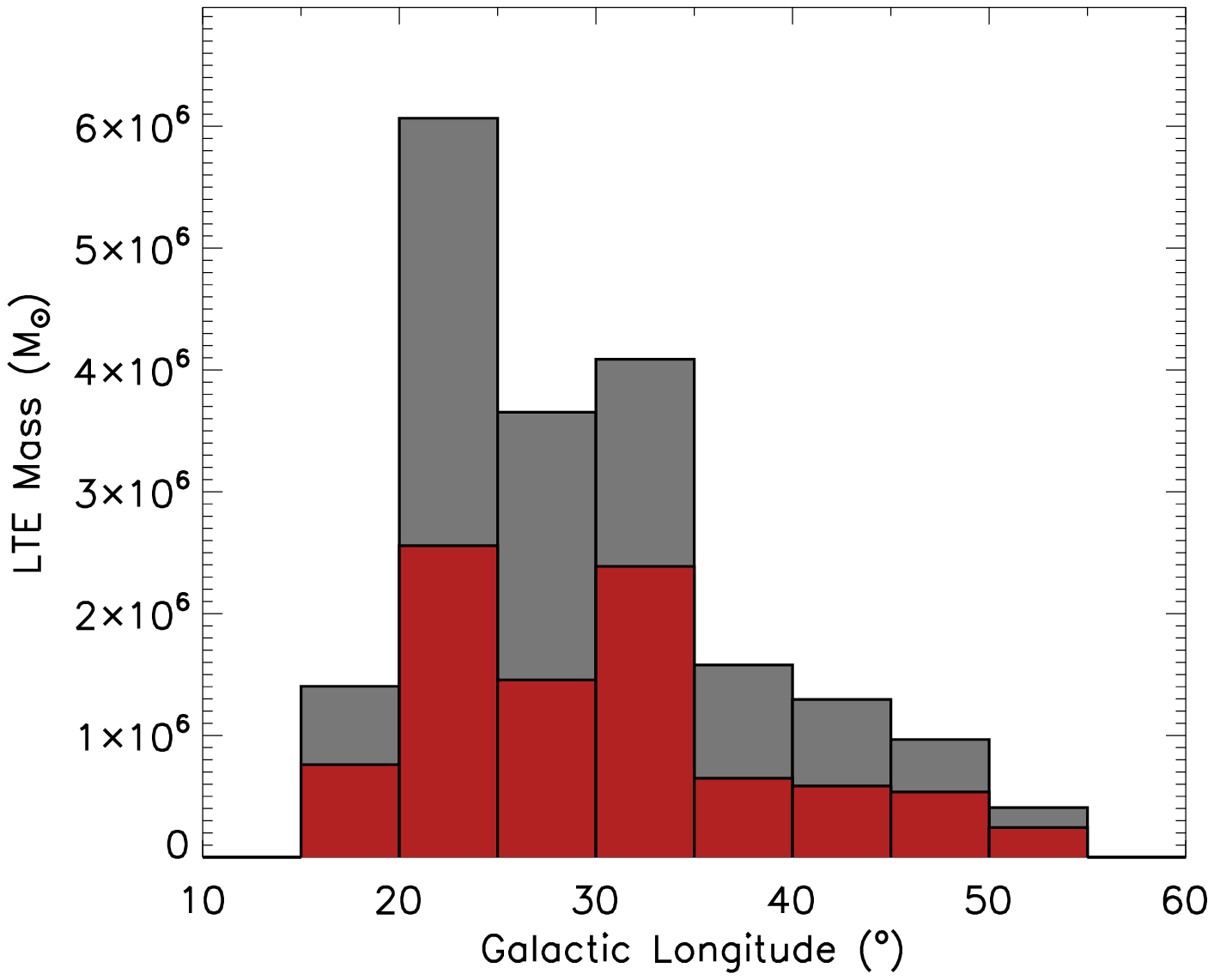}
\includegraphics[width=0.9\linewidth]{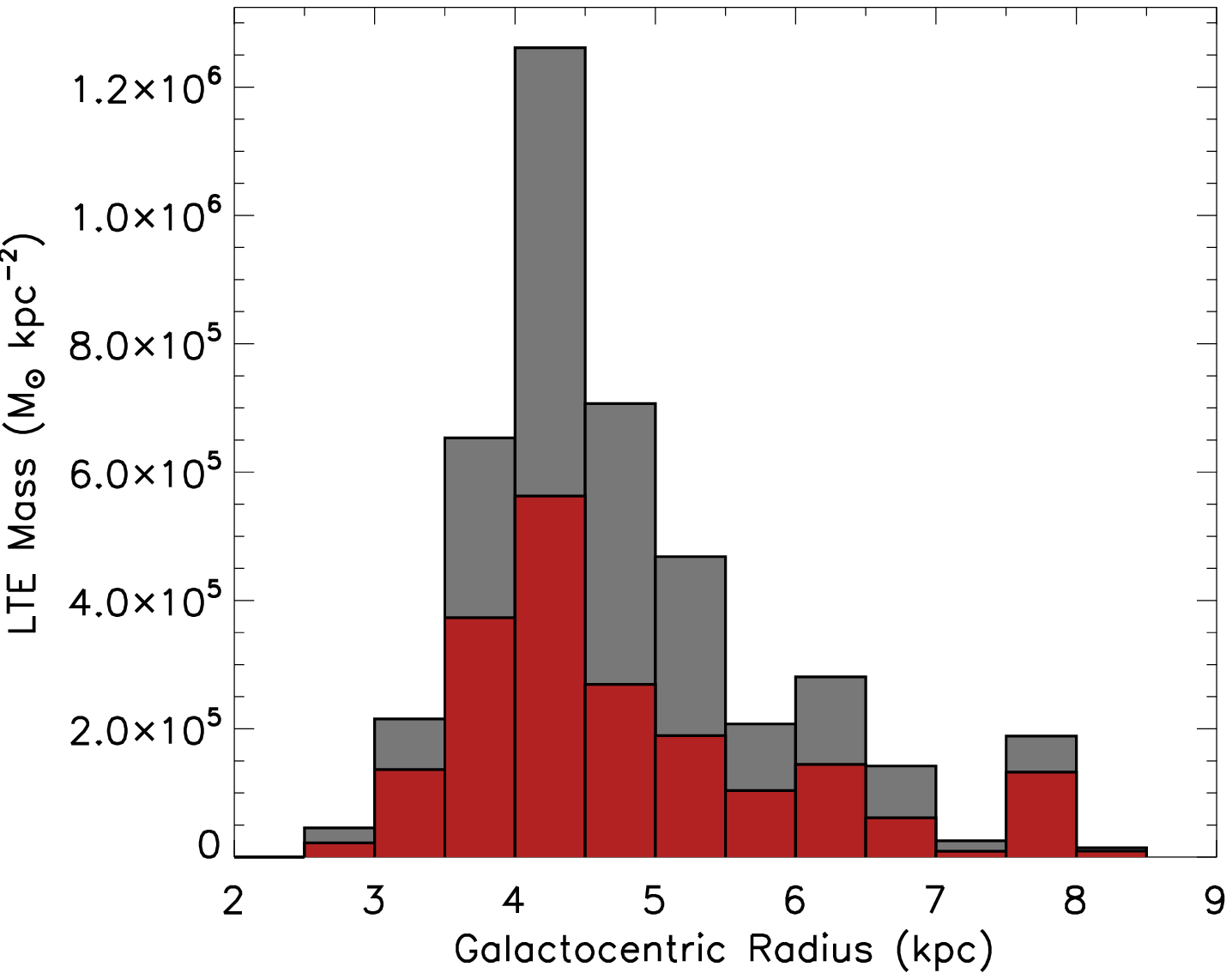}
\caption{\label{gal_mass_dist} Distribution of LTE mass as a function of Galactic longitude and mass surface density as a function of Galactocentric radius. The  whole sample of GRS clouds is shown in grey while the GRS clouds associated with an RMS source with a luminosity greater than 10$^4$\,\lsun\ are shown in red.}

\end{figure}

In this subsection we use the cloud LTE masses derived in the previous subsection to investigate the distribution of molecular material as a function of Galactic longitude and Galactocentric radius. In Fig.~\ref{gal_mass_dist} we present two histograms showing these relationships for all of the GRS clouds (grey) and the subsample of clouds associated with young massive stars above the RMS completeness limit (red). In the upper panel we show the total integrated LTE mass as a function of Galactic longitude, while in the lower panel we present the mass surface density distribution as a function of Galactocentric radius to remove any bias due to the uneven areal coverage of the GRS. 

The distribution of LTE mass as a function of Galactic Longitude (upper panel of Fig.~\ref{gal_mass_dist}) for the GRS catalogue and the subsample associated with RMS sources reveal two distinct peaks, the strongest of the two is located at $l$ $\sim$ 20--25\degr, with the second peak located at $l \sim $ 30--35\degr. The these peaks correspond with the concentration molecular gas coincident with the end of the Galactic bar and the line of sight along the tangent of the Scutum arm. These peaks also coincide with two peaks seen in the longitude distribution of the old star population reported by \citet{benjamin2005}. It is also worth noting that there is no evidence of a peak in the LTE mass distribution at the expected longitude of the Sagittarius arm tangent ($l\sim46-50$\degr). The similar absence of a peak in the distribution of the old star population was noted by \citet{benjamin2008} and has led to speculation that the Galaxy has two principle spiral arms, Scutum-Centaurus and Perseus, and two secondary or minor arms (\citealt{benjamin2008,churchwell2009}).  We will investigate this in more detail in the next section.

In the lower panel Fig.~\ref{gal_mass_dist} we present the mass surface density distribution as a function of  Galactocentric radius. This distribution reveals three peaks, the strongest is $\sim$4--4.5\,kpc, a slightly weaker peak $\sim$6--6.5\,kpc, and one at $\sim$7.5--8\,kpc. These peaks correspond to the expected galactocentric radii of the Scutum, Sagittarius and Perseus arms.

\subsection{Properties of RMS-GRS cloud associations}
\label{sect:rms_grs_clouds}

In the previous section we identified the molecular clouds involved in the formation of young massive stars by matching the RMS sources with the GRS clouds. In this section we will compare the properties of GRS clouds that are associated with the formation of the next generation of young massive stars, and those that are not associated. In the right panels of Figs.\,6 and 7 we present plots showing the distribution of RMS source luminosities and GRS cloud LTE masses as a function heliocentric distance; these plots show that we are virtually complete to young massive stars and GRS clouds above $\sim$10$^{4}$\,\lsun\ and  $\sim$10$^{3}$\,\msun, respectively. In total \citet{rathborne2009} identified 829 molecular clouds of which 595 have masses greater than $\sim$10$^{3}$\,\msun. Of these, 129 clouds are also associated with an RMS source with a luminosities greater than $\sim$10$^{4}$\,\lsun; in the following analysis we refer to these clouds as massive star forming clouds and the remaining 464 clouds as non-massive star forming clouds.

\begin{figure*}

\includegraphics[width=0.24\textwidth, trim=50 0 0 0]{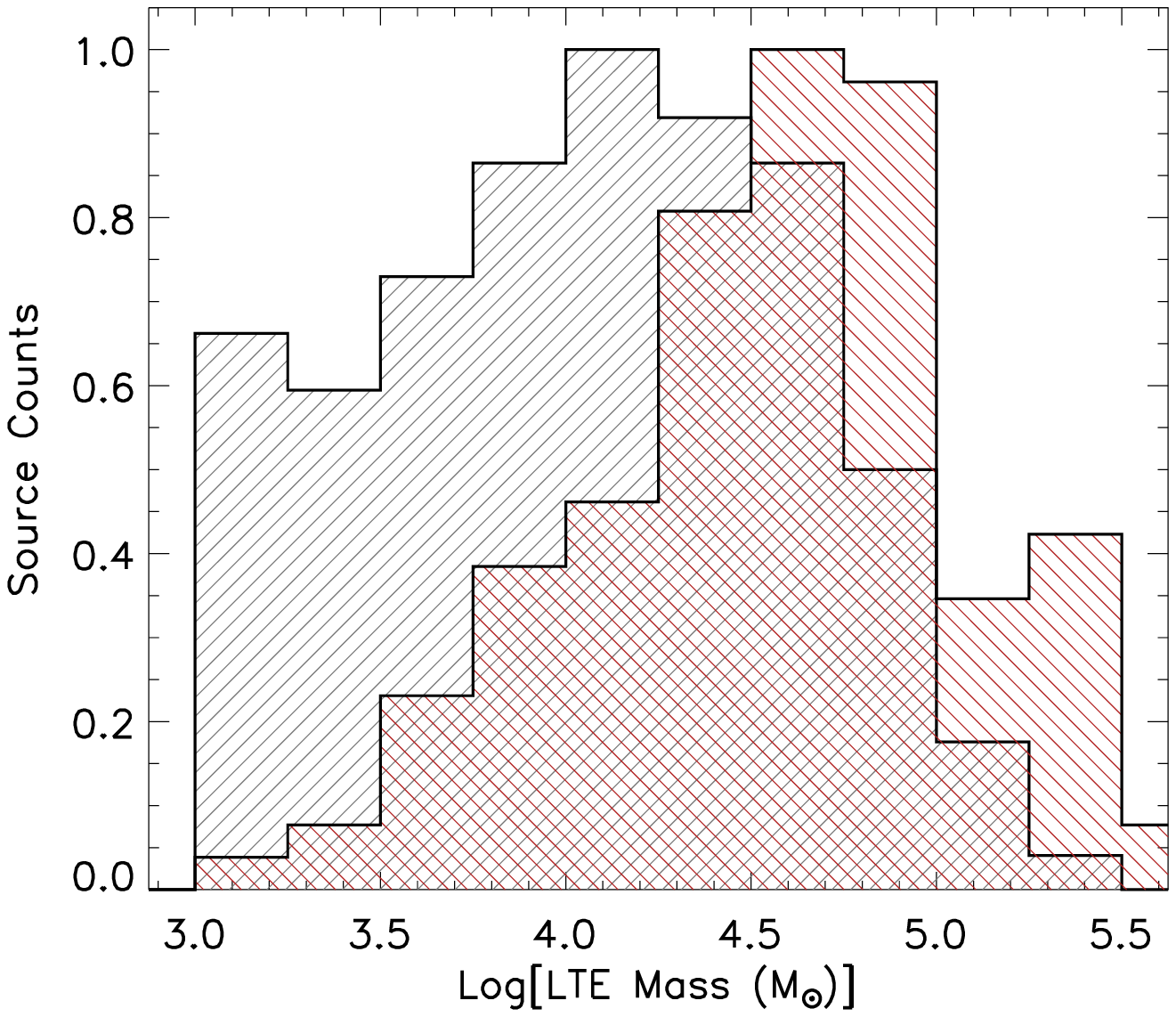}
\includegraphics[width=0.24\textwidth, trim=50 0 0 0]{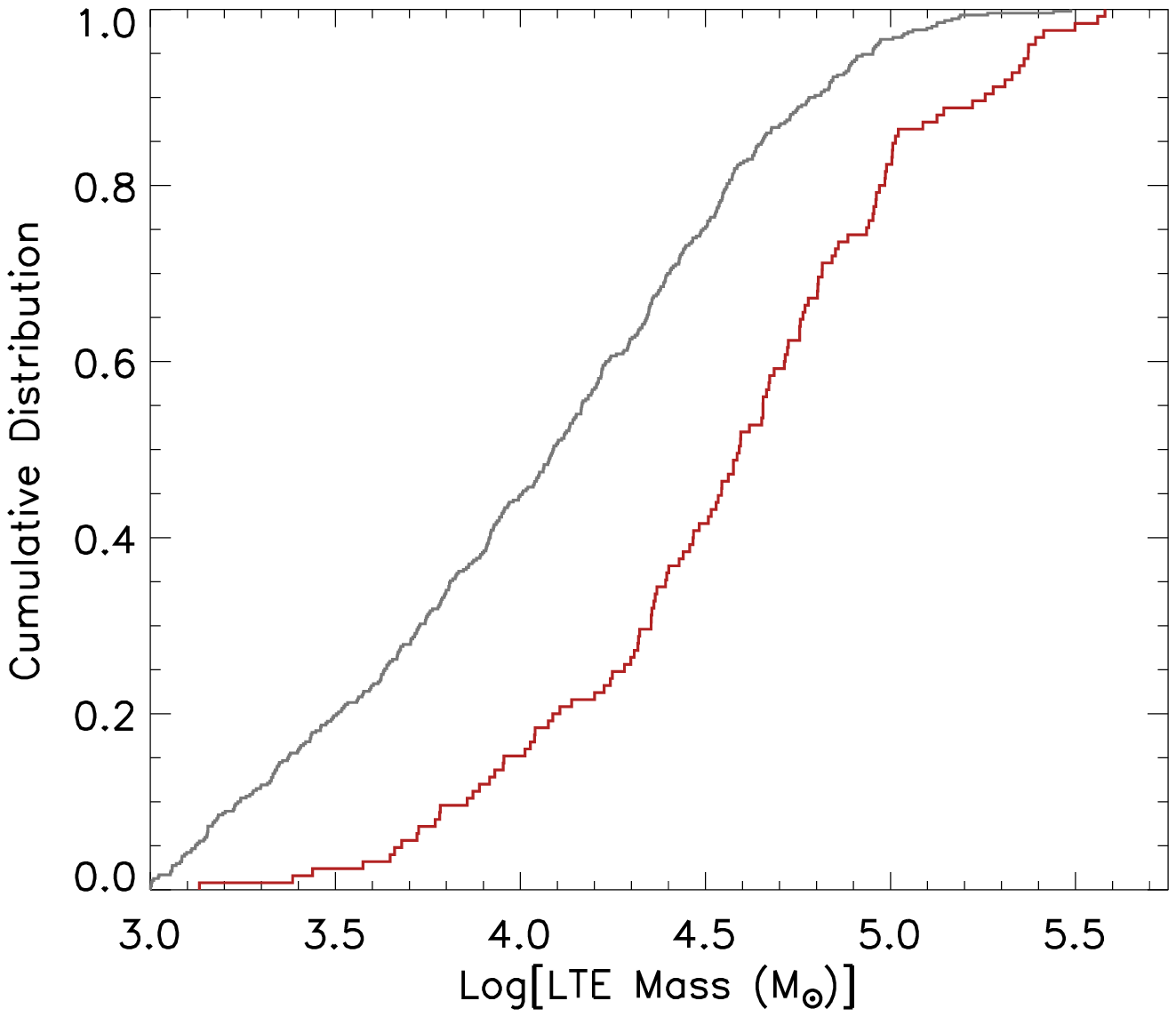}
\includegraphics[width=0.24\textwidth, trim=50 0 0 0]{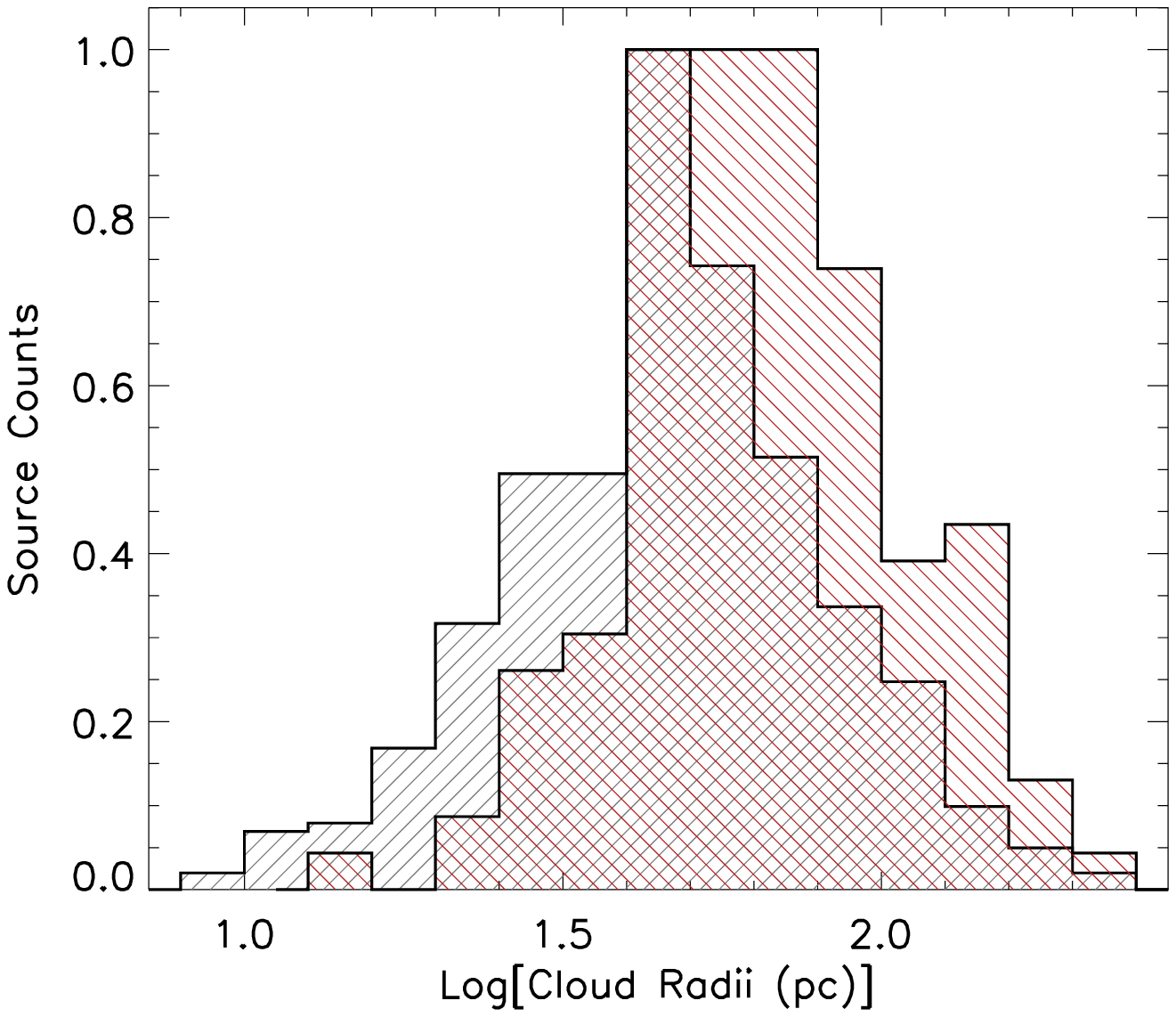}
\includegraphics[width=0.24\textwidth, trim=50 0 0 0]{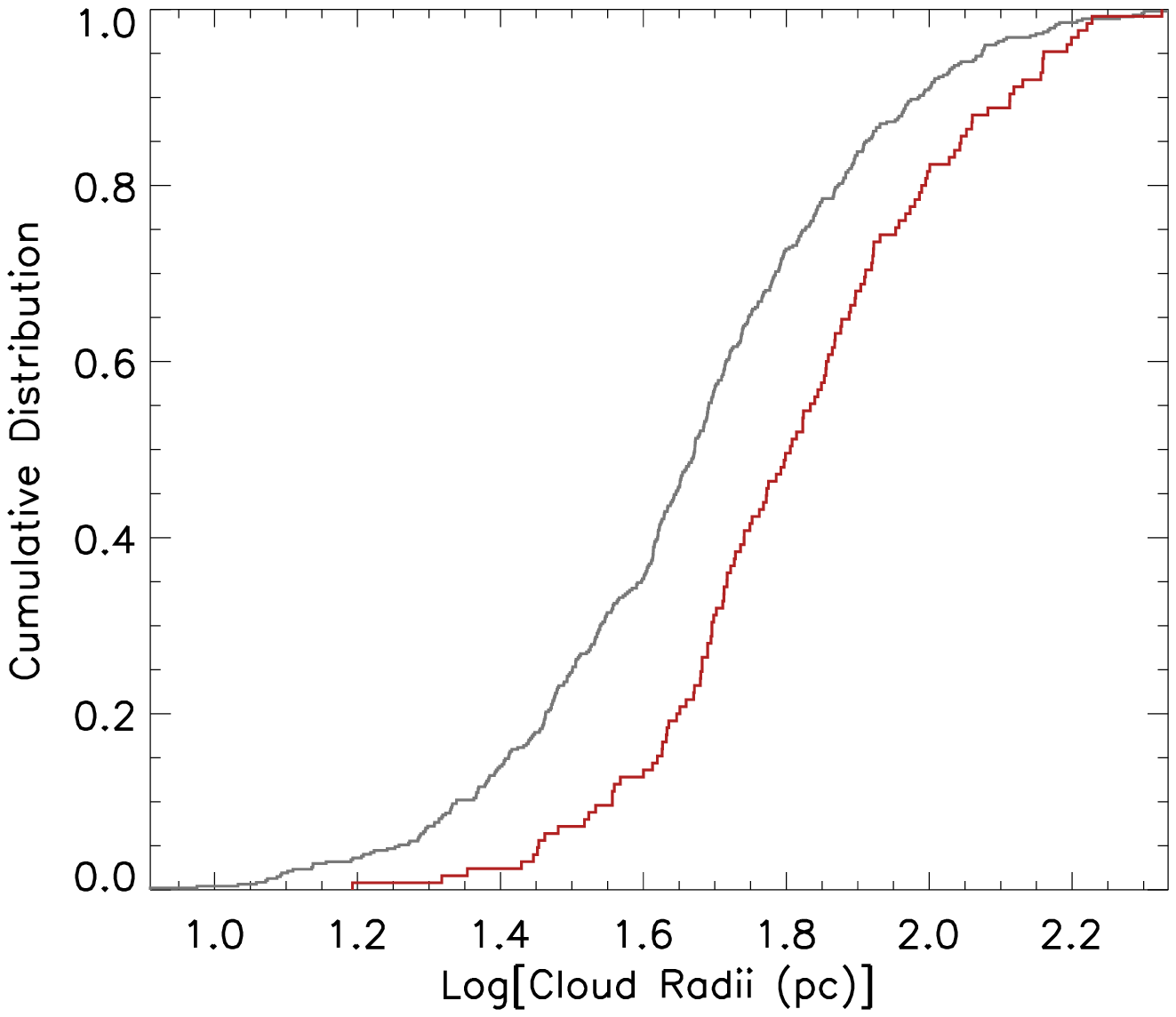}\\
\includegraphics[width=0.24\textwidth, trim=50 0 0 0]{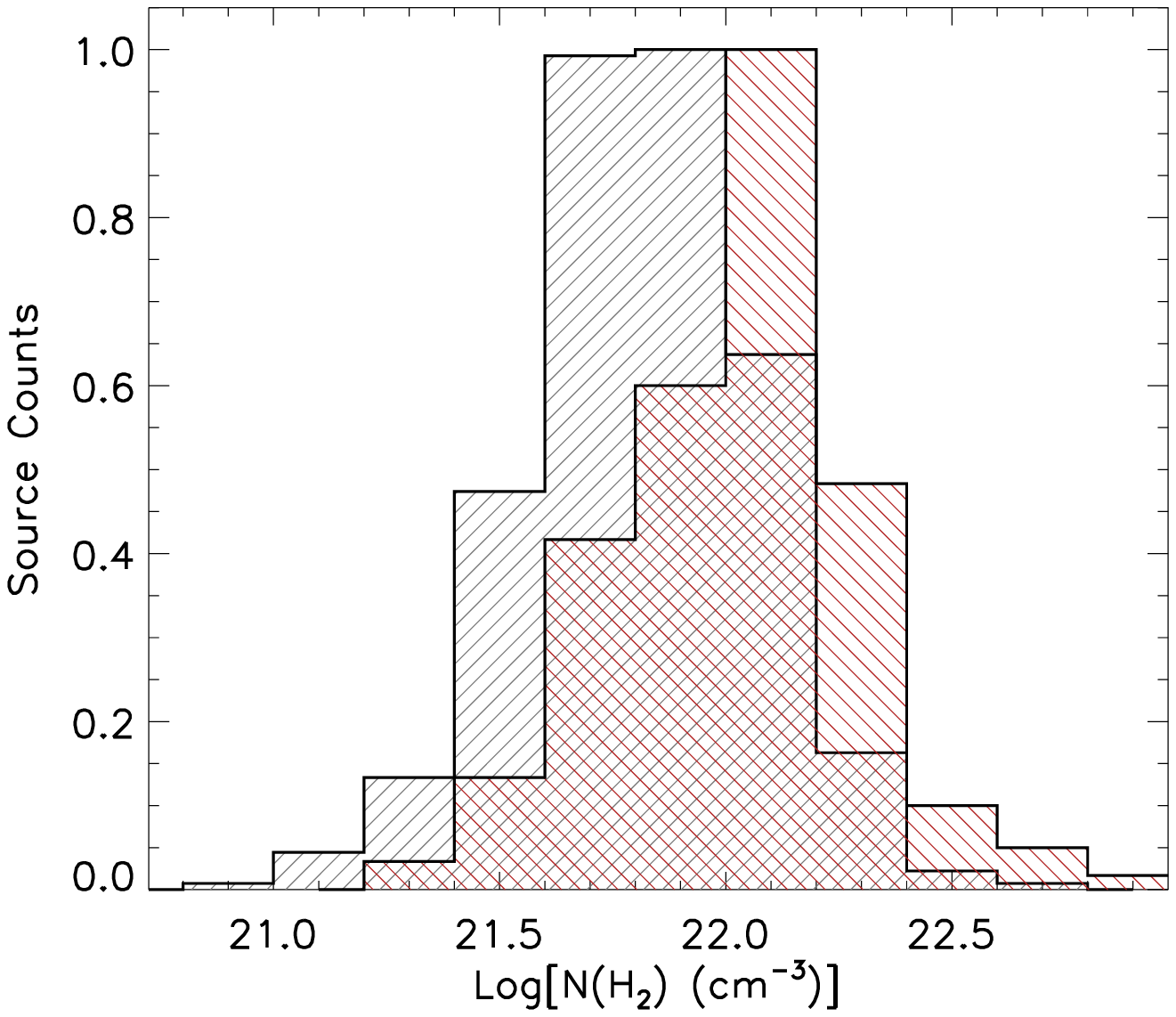}
\includegraphics[width=0.24\textwidth, trim=50 0 0 0]{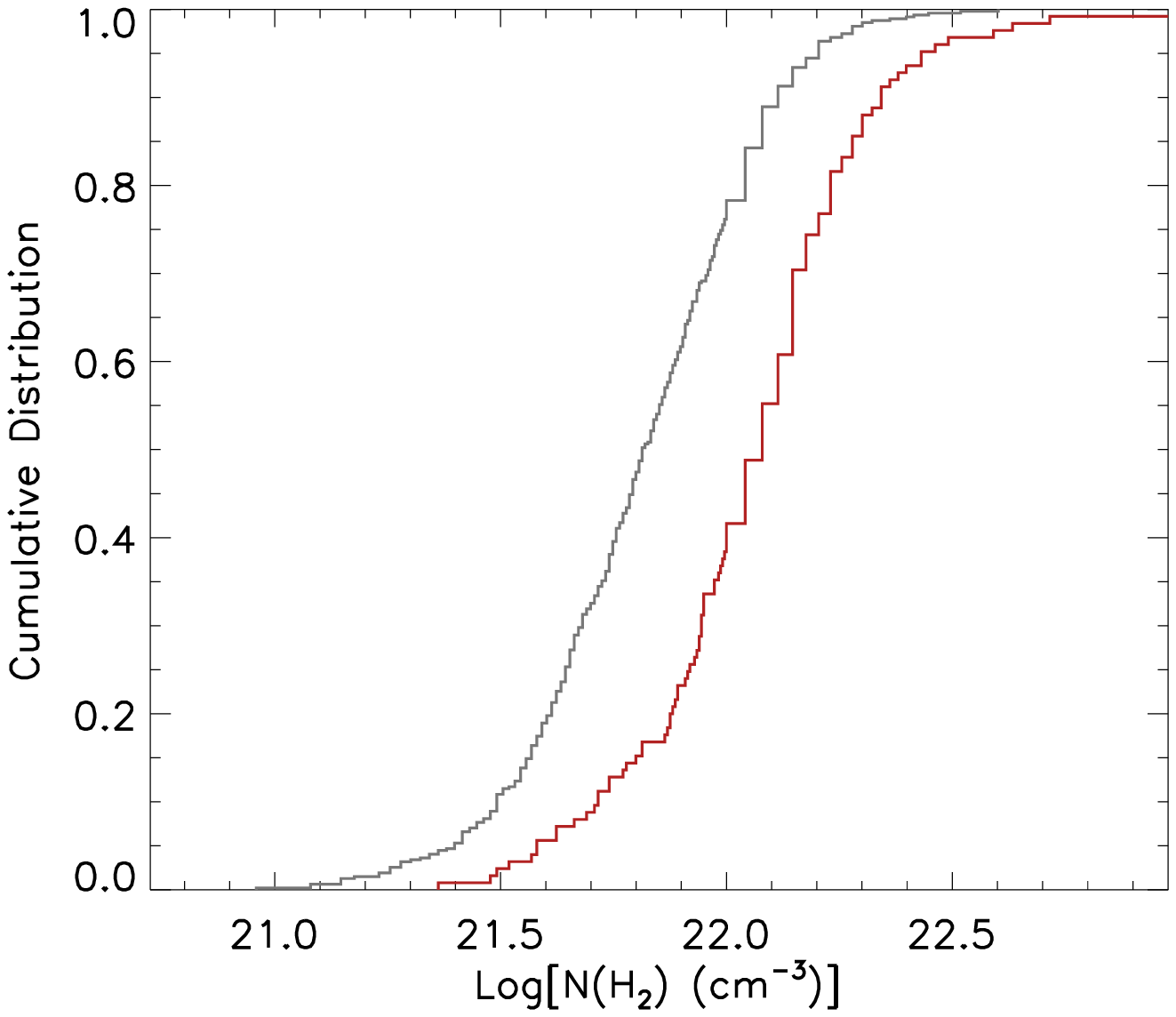}
\includegraphics[width=0.24\textwidth, trim=50 0 0 0]{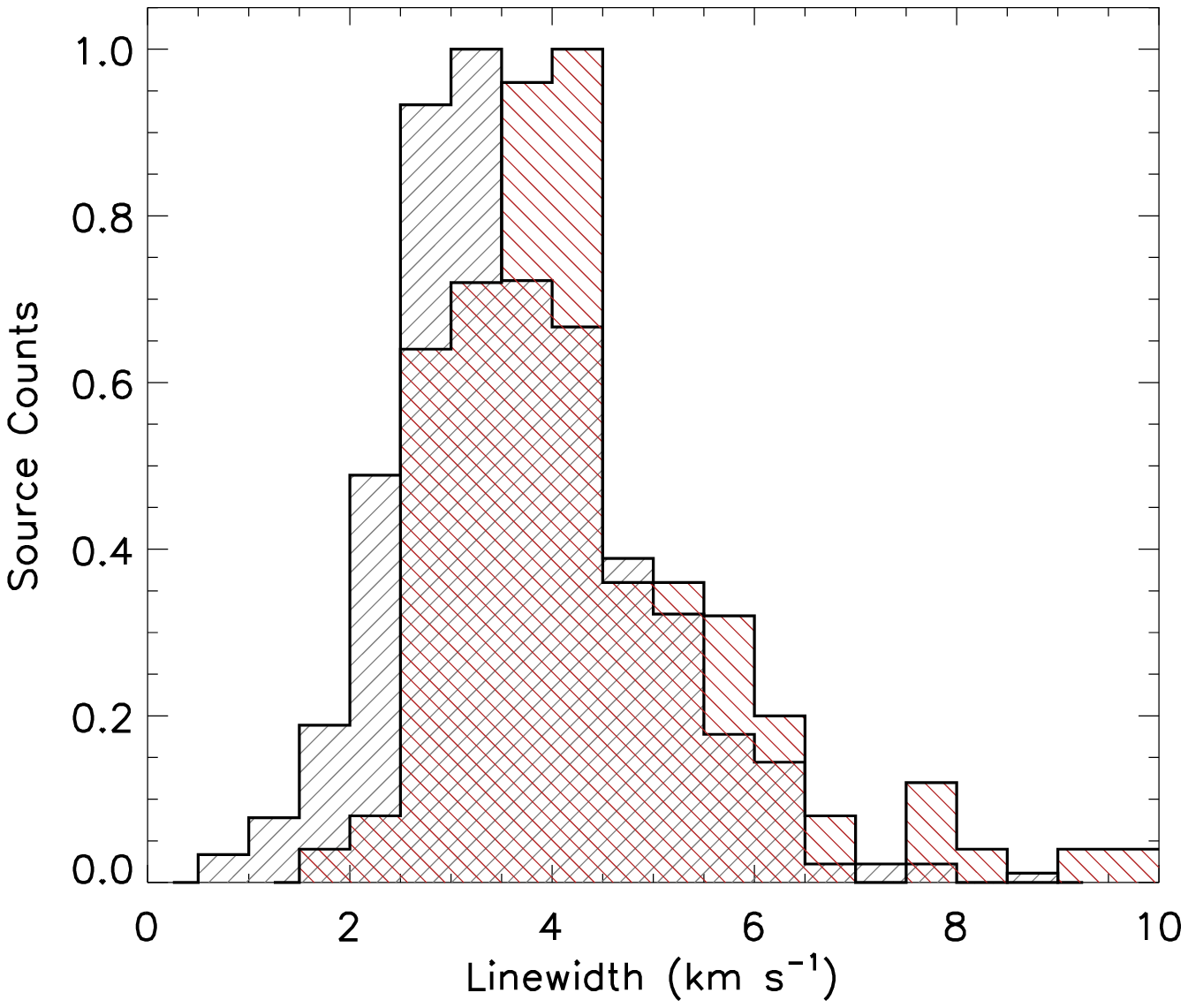}
\includegraphics[width=0.24\textwidth, trim=50 0 0 0]{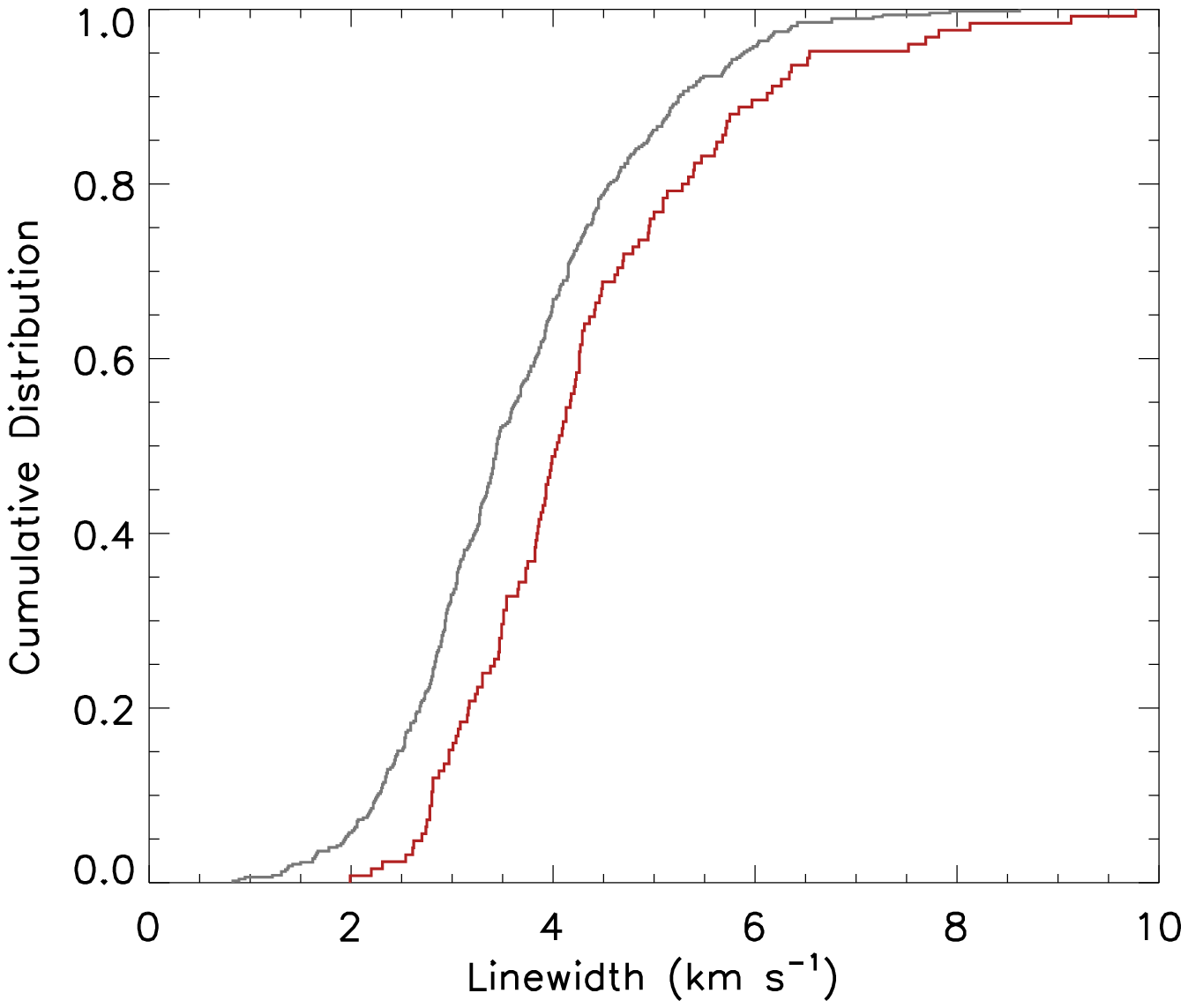}\\
\includegraphics[width=0.24\textwidth, trim=50 0 0 0]{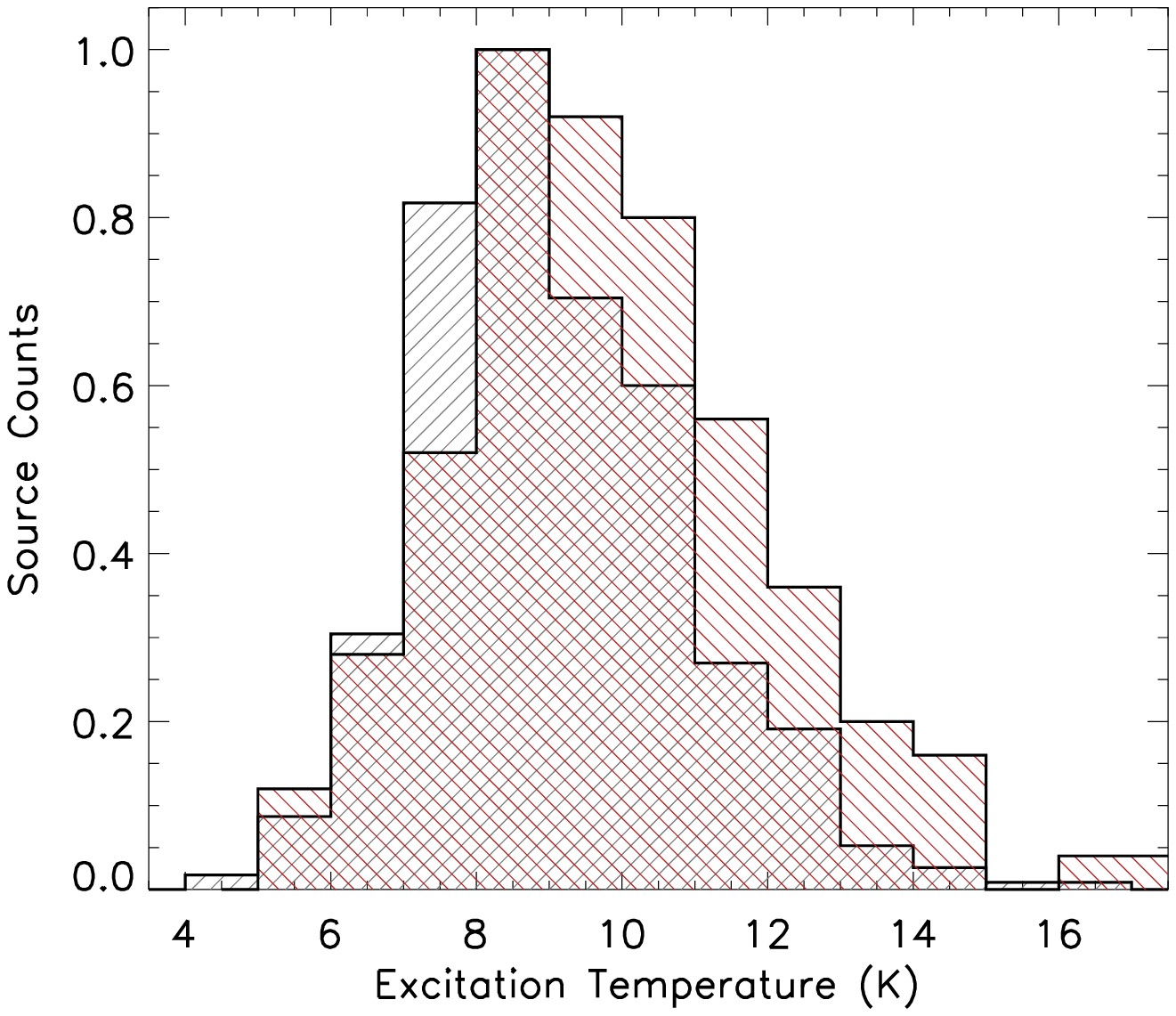}
\includegraphics[width=0.24\textwidth, trim=50 0 0 0]{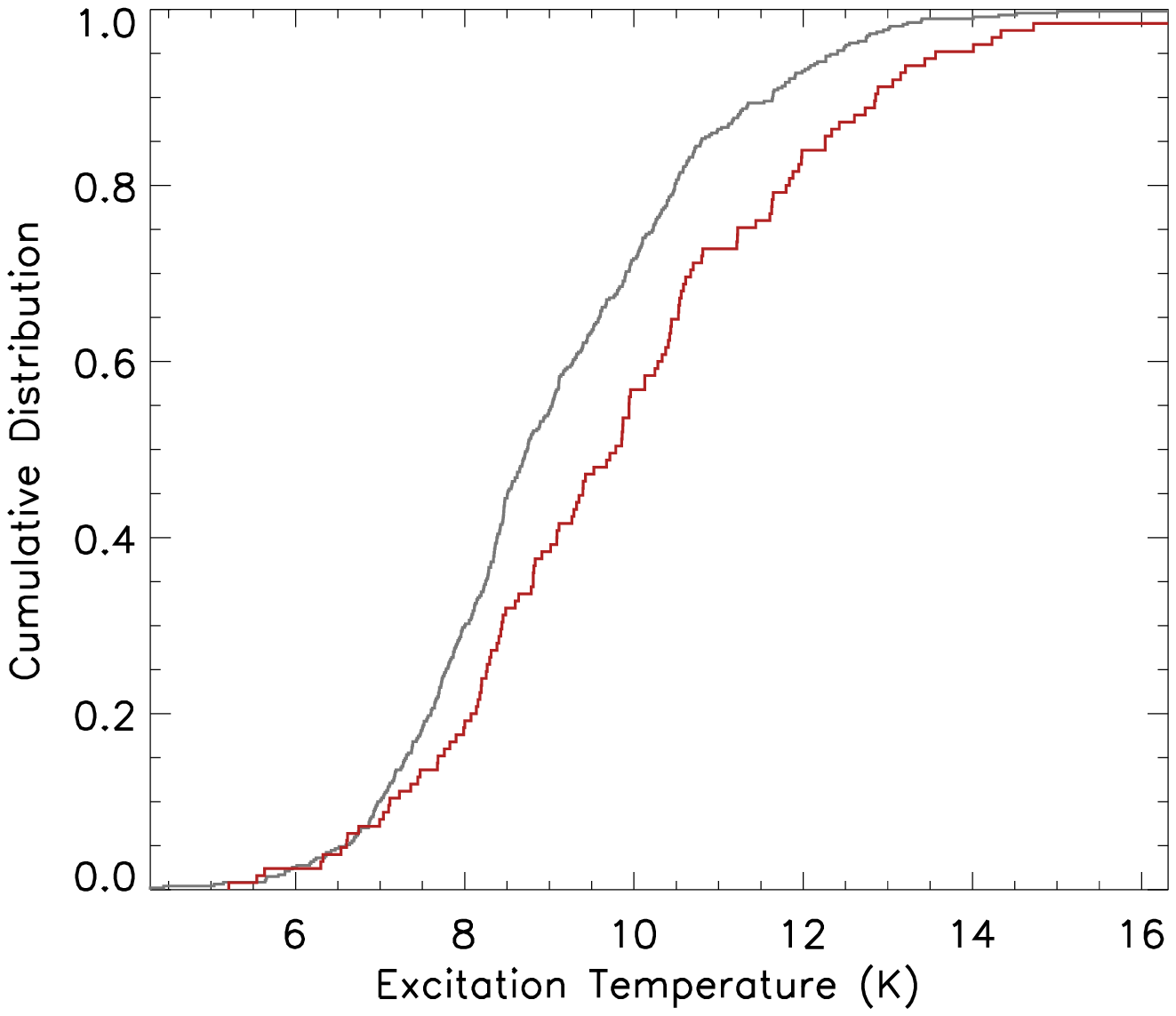}
\includegraphics[width=0.24\textwidth, trim=50 0 0 0]{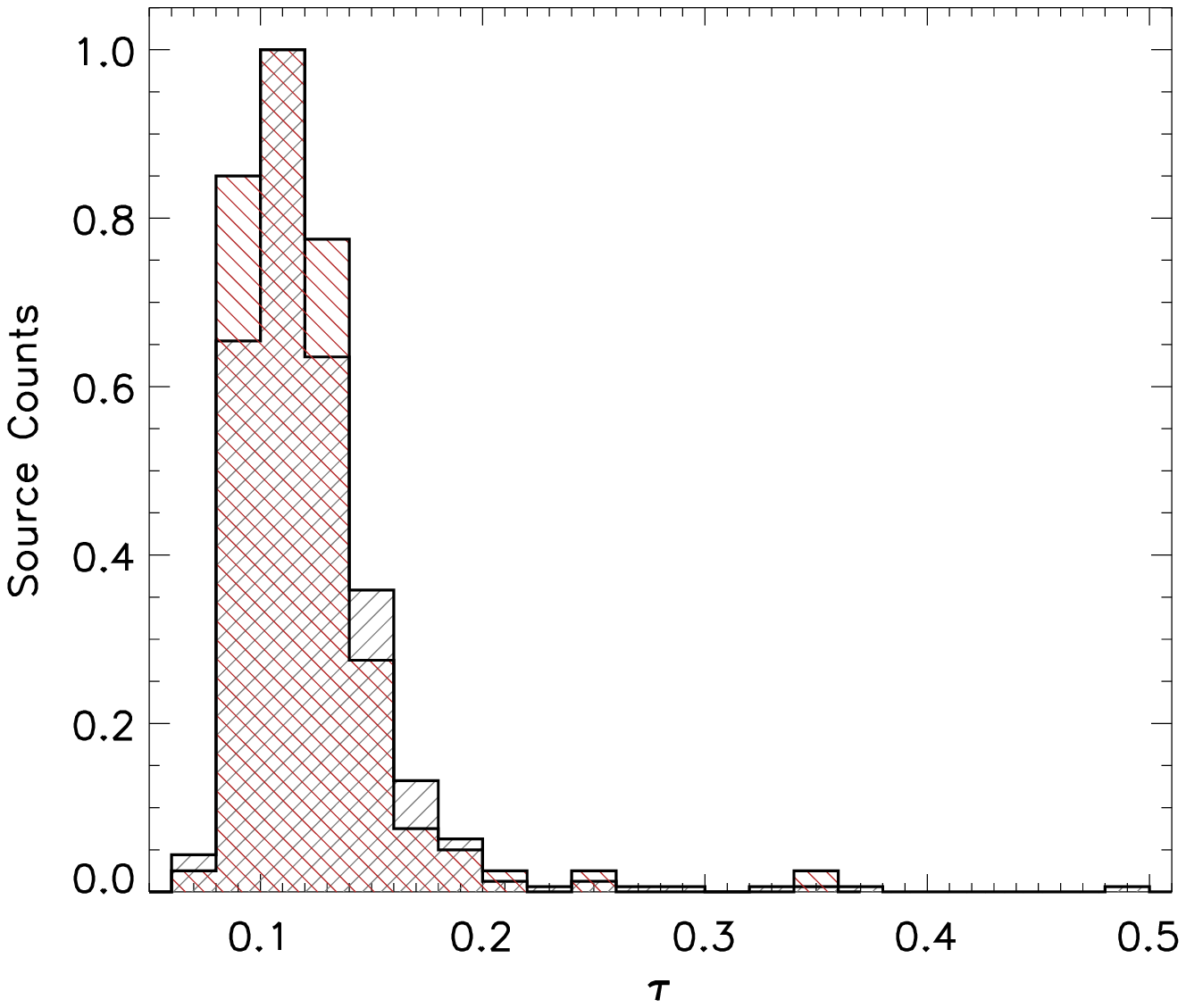}
\includegraphics[width=0.24\textwidth, trim=50 0 0 0]{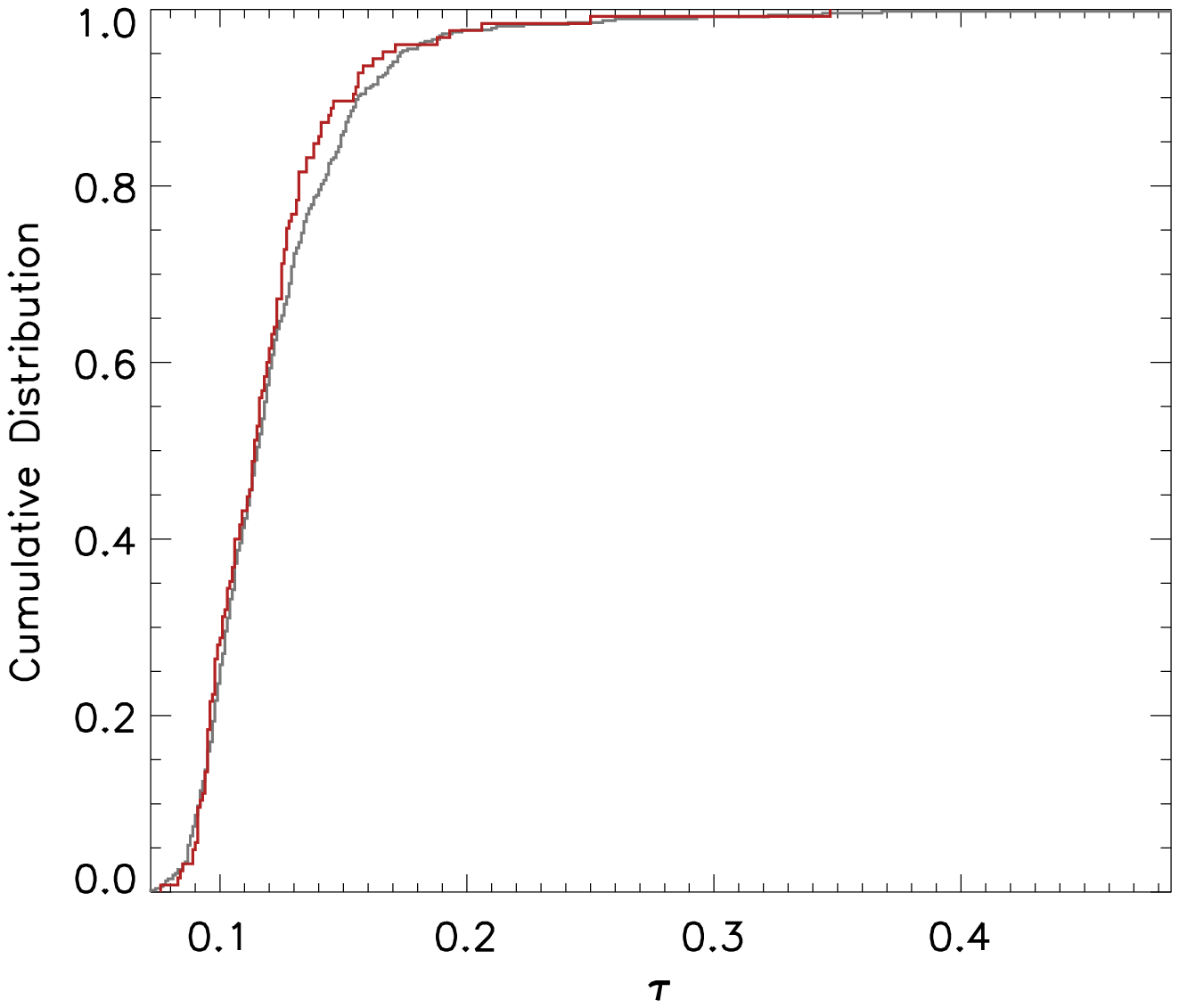}\\

\caption{\label{fig:cloud_properties} Histograms for a selection of derived parameters chosen to show the distribution of physical parameters of clouds associated with the formation of massive stars (red) and the non-massive star forming clouds (grey).}
\end{figure*}

\begin{table*}
\begin{center}
\caption{Summary of general parameters derived from the GRS clouds with masses above 10$^3$\,\msun. We have separated this sample of clouds into two groups; those found to be associated with a young massive star above the RMS completeness limit (i.e., $L_{\rm{Bol.}}\geq10^4$\,\lsun) are classified as massive star forming with all other clouds being classified as non-massive star forming clouds.}
\label{tbl:cloud_comparions}
\begin{minipage}{\linewidth}
\begin{tabular}{l...........c}
\hline
\hline
& \multicolumn{5}{c}{Non-Massive Star Forming Clouds} && \multicolumn{5}{c}{Massive Star Forming Clouds}\\
\cline{2-6}\cline{8-12}
\multicolumn{1}{l}{Parameter}	&	\multicolumn{1}{c}{Min} & \multicolumn{1}{c}{Max} &\multicolumn{1}{c}{Mean}	&\multicolumn{1}{c}{Median} &	\multicolumn{1}{c}{Std. dev.}	&&	\multicolumn{1}{c}{Min} & \multicolumn{1}{c}{Max} &\multicolumn{1}{c}{Mean}	&\multicolumn{1}{c}{Median} &	\multicolumn{1}{c}{Std. dev.}	& \multicolumn{1}{c}{K-S test} \\
\hline
Log[LTE Mass ($M_\odot$)] 		& 3.0 		& 5.4 	&4.1	&4.1& 	0.6		&& 3.1 &5.6	&4.6 &4.6 			& 0.5	& $\ll0.01$	\\

Log[Cloud Radii (pc)] 		& 0.9 		& 2.3 	&1.7	&1.7& 	0.2		&& 1.2 &2.3	&1.8 &1.8 			& 0.2	& $\ll0.01$	\\

Log[Peak H$_2$ col. den. (cm$^{-2}$)] 		 &21.0&22.6&21.8&21.8& 0.3		&& 	21.4& 23.0& 	22.1	&22.1 &  	0.3	& 	$\ll0.01$	\\

FWHM \vlsr\ (km s$^{-1}$)		& 0.8 		& 8.6 	&3.6	&3.4& 	1.2		&& 2.0 &9.8	&4.3 &4.0 			& 1.4	& $\ll0.01$	\\

Excitation Temperature (K)		& 4.3 		& 16.3 	&9.1	&8.7& 	1.8		&& 5.2 &17.4	&9.9 &9.8 			& 2.3	& $\ll0.01$	\\

$\tau$		& 0.07 		& 0.49 	&0.12	&0.12& 	0.04		&& 0.08 &0.34	&0.12 &0.11 			& 0.03	& 0.48	\\

\hline
\end{tabular}\\

\end{minipage}
\end{center}

\end{table*}

In Fig.\,\ref{fig:cloud_properties} we present normalised histograms and cumulative distribution function (CDF) plots for a number of physical parameters for both the massive star forming clouds (red) and the non-massive star forming clouds (grey). These histograms and CDF plots clearly illustrate the significant differences between the two samples of clouds. 

We used Kolmorgorov-Smirnov (K-S) tests to evaluate the significance of the parametric differences between the two cloud samples. We are able to reject the null hypothesis that the two population are drawn from the same parent population with confidence values above 3$\sigma$ for all of the parameters mentioned in the previous paragraph. The only parameter that is not significantly different between the two cloud samples is the optical depth ($\tau$). In Table~\ref{tbl:cloud_comparions} we present a summary of the averages, minimum and maximum values and the result of Kolmorgorov-Smirnov (K-S) tests comparing the massive star forming and non-massive star forming clouds.

The clouds associated with massive star formations are in generally larger, more massive and, perhaps not suprisingly, are found to have significantly higher column densities. Furthermore, the massive star forming clouds are warmer and more turbulent as indicated by the higher excitation temperatures and larger velocity dispersion than found towards the non-massive star forming clouds. The only parameter that is not significantly different between the two cloud samples is the optical depth ($\tau$).

\section{Galactic distribution of young OB stars}
\label{sect:discussion}

In this section we will use the distance and luminosity results obtained previously to investigate the spatial distribution of massive stars in the Galaxy. Regions of massive star formation are almost exclusively found to be associated with the spiral arms where molecular clouds are thought to form (\citealt{kennicutt2005}). The Galactic distribution of massive young stars is, therefore, an important probe of Galactic structure.

\begin{figure*}
\includegraphics[width=0.49\linewidth, trim= 50 0 50 0]{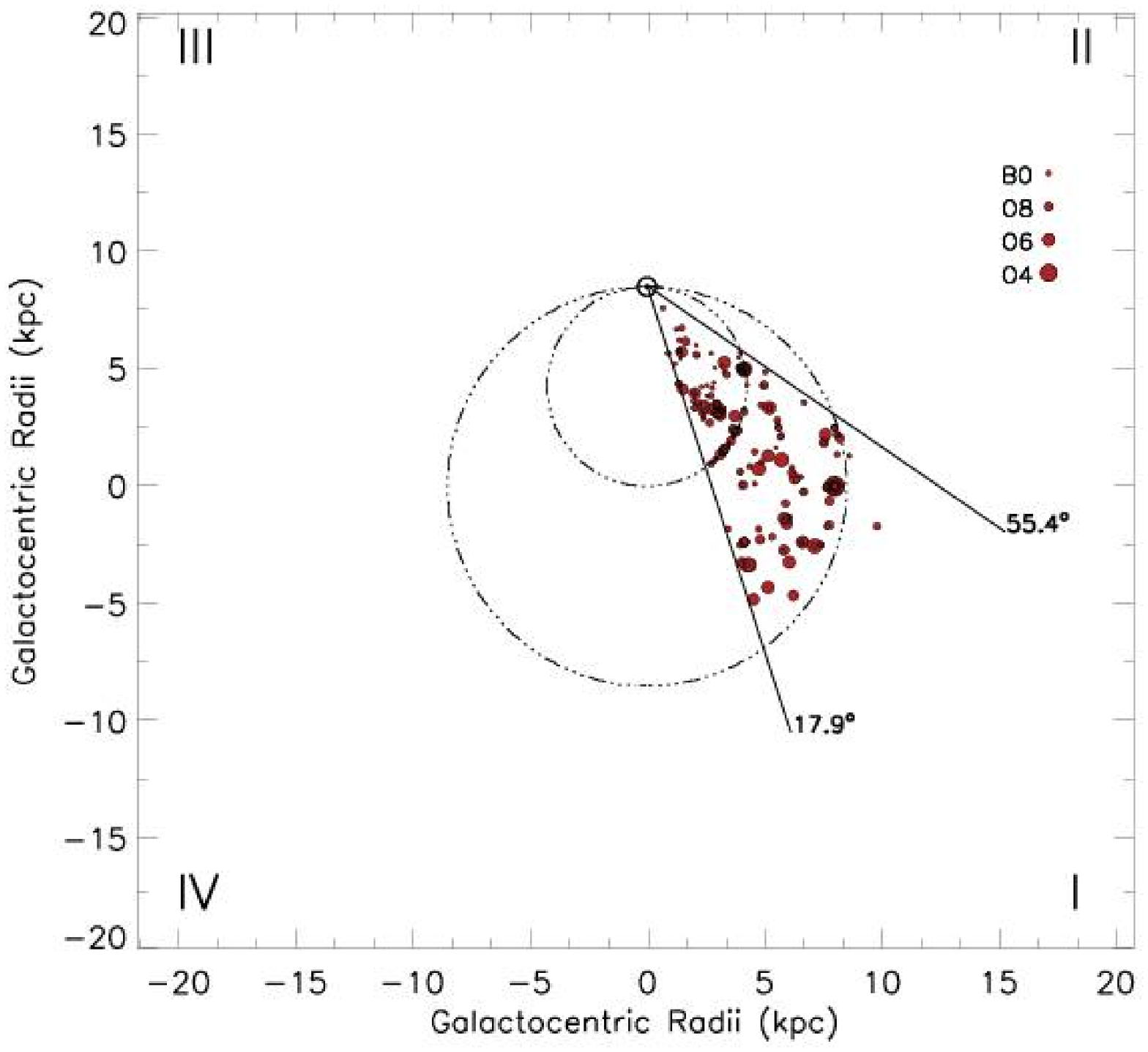}
\includegraphics[width=0.49\linewidth, trim= 50 0 50 0]{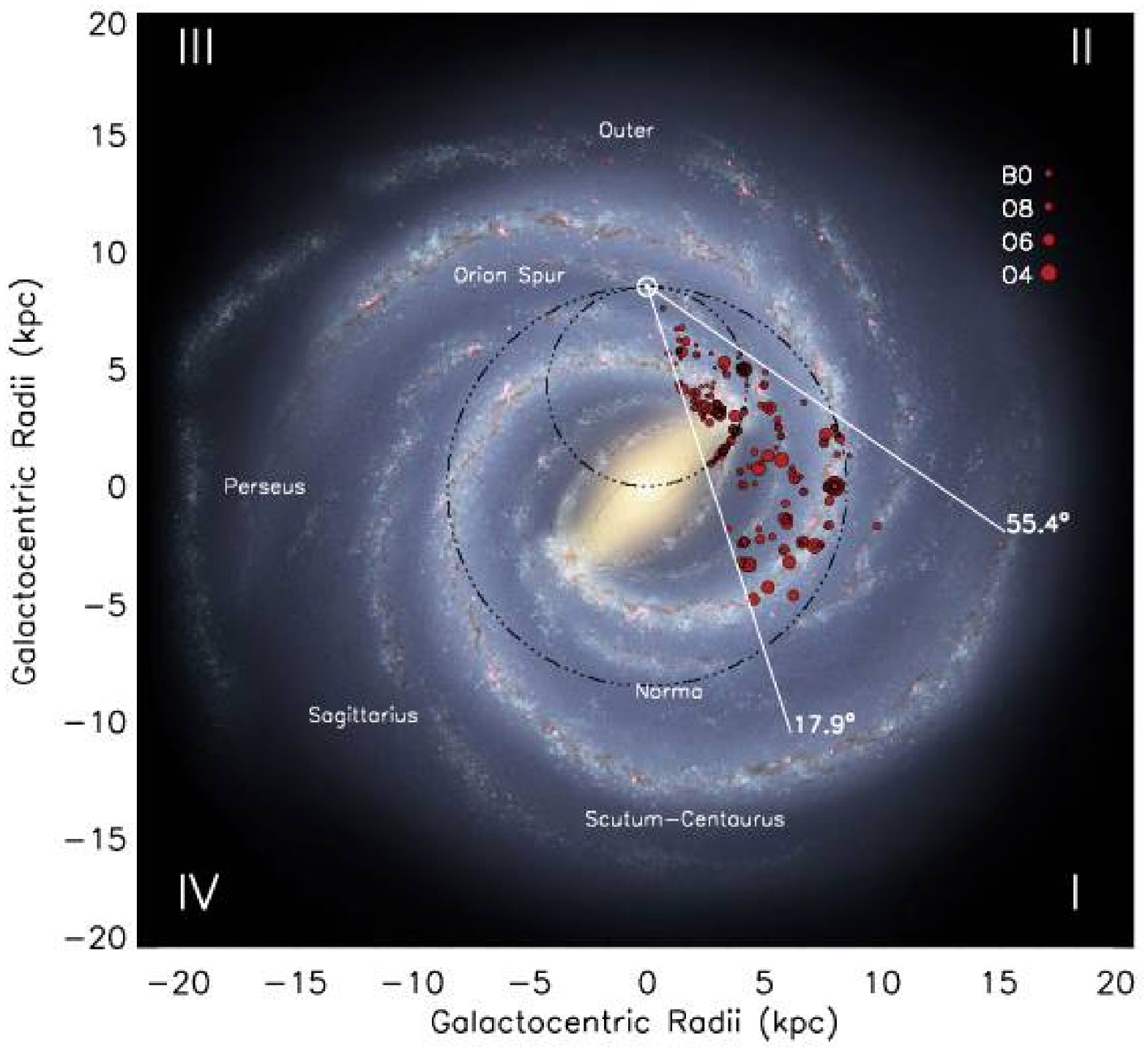}

\caption{\label{fig:gal_dist} Galactic distribution of the complete RMS sample of MYSOs and UC\,HII regions with bolometric luminosites $\geq$ 10$^4$\,\lsun in the GRS field. In both panels we show the kinematic positions of our sample as red-in-black circles, the sizes of which give an indication of their respective luminosities. In the upper right corner of each plot we give the luminosities for a sample of zero age main sequence stars. In the left panel we plot the positions of our sample without any Galactic structure information to lead the eye. In the right panel we have superimposed the RMS source distribution over a sketch of how the Galaxy is thought to appear if viewed face-on. This image has been produced by Robert Hurt of the Spitzer Science Center in consultation with Robert Benjamin and attempts to synthesise many of the key elements of Galactic structure using the best data currently available (see text for more details). The position of the Sun is shown by the small circle above the image centre. The Roman numerals in the corners refer to the Galactic quadrants and the two thick lines originating from the location of the Sun enclose the region of the Galactic Plane covered by the GRS.  The dot-dashed circles represent the locus of tangent points and the Solar Circle.}

\end{figure*}

In Fig.\,\ref{fig:gal_dist} we present two plots showing the Galactic distribution of our complete sample of young massive stars. In both plots the positions of the RMS sources ($L_{\rm{Bol.}}\geq$ 10$^4$\,\lsun) are indicated by filled red circles, the sizes of which provides an indication of their bolometric luminosities. For RMS sources that have been associated with a molecular cloud we have assumed the RMS source distance is the same as its host cloud. Using the systemic velocity of the cloud smooths out localised velocity perturbations that might lead to larger scatter in the distances. Distances for the high latitude sources (i.e., $|b|> 1\degr$) have been taken from \cite{urquhart_13co_north}.

The left panel of this figure simply presents the positions of the sample without the addition of Galactic features to lead the eye. Examining the distribution shown in this plot there are no immediately obvious structures, however, one can begin to see structures that suggest the presence of spiral arms between the locus of tangent points and the solar circle.

In the right panel of Fig.\,\ref{fig:gal_dist} we plot the positions of the RMS sources over an image of the Galaxy produced by Robert Hurt of the Spitzer Science Center in consultation with Robert Benjamin (University of Wisconsin-Whitewater). This image attempts to synthesise all that has been learnt about Galactic structure over the past fifty years including: a 3.1-3.5\,kpc Galactic Bar at an angle of 20\degr\ with respect to the Galactic Centre-sun axis (\citealt{binney1991,blitz1991,dwek1995}), a second non-axisymmetric structure referred to as the ``Long Bar'' (\citealt{hammersley2000}) with a Galactic radius of $4.4\pm0.5$\,kpc at an angle of 44\degr\ $\pm 10$\degr\ \citep{benjamin2005}, the Near and Far 3-kpc arms, and the four principle arms: Norma, Sagittarius, Perseus and Scutum-Centaurus. The position of the arms is based on the \citet{georgelin1976} model which has been modified to incorporate Very Long Baseline Array maser parallax measurements (e.g., \citealt{xu2006}) and refined directions for the spiral arm tangents from \citet{dame2001}. The Perseus and Scutum-Centaurus arms have been emphasised in this image to reflect the overdensities seen in the old stellar disk population towards their expected Galactic longitudes tangent positions \citep{benjamin2008,churchwell2009}. 

Comparing the distribution of our sample of young massive stars with the arms shown in the right panel of Fig.\,\ref{fig:gal_dist} reveals them to be strongly correlated. The highest density of massive stars is coincident with the end of the Galactic bar and the proposed location of the start of the Scutum-Centaurus arm. The high concentration of young massive stars towards the end of the bar also correlates with the strongest peak in the gas mass distribution (as traced by the $^{13}$CO luminosity; see \citealt{roman2009} for more details). Both distributions peak approximately at the leading edge of the bar, falling off steeply in the direction of rotation, and tailing off behind the bar. This high density of massive stars is also coincident with the co-rotation radius ($\sim$4.5\,kpc). It is therefore unclear whether the high density of massive stars is the result of dynamical interaction between the bar and the spiral arms or co-rotation.

\begin{figure}

\includegraphics[width=0.95\linewidth, trim= 20 0 0 20]{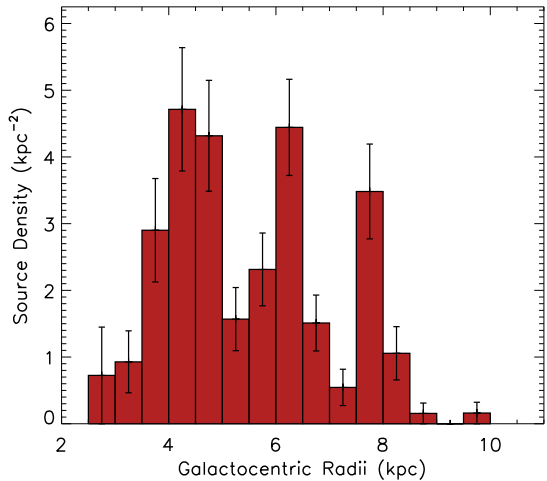}

\caption{\label{fig:gal_rgc_dist} Histogram showing the surface density distribution of young massive stars as a function of Galactocentric radius. Errors have been calculated assuming \poi\ statistics (i.e., $\sqrt{N}$ where $N$ is the number of sources in each bin).
}

\end{figure}

We also note the correlation between the massive young stars and the Sagittarius and Perseus arms, this is particularly strong towards the Perseus arm. We note that the correlation of RMS sources with the far section of the Sagittarius arm is not as strong as seen towards the near section, with a number of sources lying between the Sagittarius and Perseus arms in the model. This may suggest that the model needs small adjustments or that we have a small systematic error in these sources. Comparing the position of the tangent-circle sources and the far 3\,kpc arm we see correlation between  them, which suggests they are associated. In Fig.\,\ref{fig:gal_rgc_dist} we show the source density per kpc$^{-2}$ of young massive stars as a function of Galactocentric radius. This plot clearly reveals three significant peaks at approximately 4, 6 and 8\,kpc. Inspection of the spatial distribution plot (Fig.\, 9) shows that the strongest peak at 4\,kpc coincides with the intersection of the Long Bar and the Scutum-Centaurus arm, and that the second and third peaks are at the Galactocentric radii of the Sagittarius and Perseus arms, respectively. The distribution with Galactocentric radius is only dependent on the Galactic rotation curve and not on the solution of near-far ambiguity. Therefore, the correlation of these peaks with the structures in the spiral-arm model lends support to their existence.

A similar Galactocentric radial distribution is seen in the $^{13}$CO data presented in Fig.\,7 and has been reported for $^{12}$CO by \citet{clemens1988} and for a large sample of HII regions by \citet{anderson2009a}, however, in the case of the HII regions they do not find a peak at $\sim$ 8\,kpc. It is not clear why there is a peak in the distribution of young massive stars at this distance from the Galactic centre which is not seen in the \citet{anderson2009a} sample of HII regions. However, there is evidence of streaming motions in HI data at all three Galactocentric radii, which are a strong indication these are the locations of spiral arms (\citealt{mcclure2007}).

In a recent study of IRDCs (\citealt{jackson2008}) reported finding a significant concentration of sources at $\sim$8\,kpc in the 1st Quadrant. Given than the identification of IRDCs is biased towards more nearby clouds the authors assumed their sources were all located at the near kinematic distance, and thus, the observed source enhancement was due to a high density of sources in the solar neighbourhood. \citet{jackson2008} consider the possibility that the peak could be due to sources located at the far side of the Galaxy, possibly associated with a distant spiral arm, but conclude that this is less likely. We can rule out any possibility that the peak found at $\sim$ 8\,kpc in our sample is due to high concentrations of sources located in the vicinity of the Sun since the luminosity cut has filtered out nearby low- and intermediate-mass objects. The correlation of the $\sim$8\,kpc peaks in both the IRDCs and our sample of massive stars suggests that the peak of IRDCs seen at 8\,kpc may in part be due to a high concentration of IRDCs associated with the Perseus arm at the far side of the Galaxy. 

The positional coincidence of a large number of young massive stars with the proposed location of the Sagittarius arm seen in Figs.\,9 and 10 provide convincing evidence that a significant amount of star formation is taking place within this arm. Inspection of the source density would suggest there are similar levels of star formation associated with all three spiral arms in this Galactic quadrant. These results are consistent with a model of the Galaxy consisting of four principle arms. However, these results appear to be in contradiction with the results of the 4.5\,\mum\ number counts (\citealt{benjamin2005}) that trace the longitude distribution of the old stellar disc populations; these data fail show any evidence of an overdensity in the old stellar disc population at the expected tangent direction of the Sagittarius arm ($l=46$--$50$\degr). 

\begin{figure}

\includegraphics[width=0.95\linewidth, trim= 20 0 0 20]{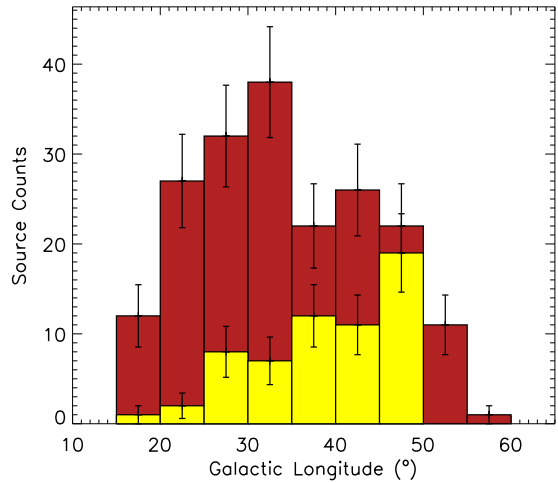}

\caption{\label{fig:gal_long_dist} Distribution of RMS sources as a function of Galactic longitude. The histogram of the whole sample of young massive stars is shown in red with sources associated with the Sagittarius arm shown in yellow. Errors have been calculated assuming \poi\ statistics.
}

\end{figure}

The tangent direction of the Sagittarius arm lies only a couple of kpc in front of the Perseus arm and it is possible that the overdensity is being masked by a high stellar background associated with the Perseus arm. This is impossible to test using the old stellar disk population since there is no velocity data available. However, we can test this hypothesis using the RMS sample. In Fig.\,\ref{fig:gal_long_dist} we present a plot of the source density as a function of Galactic longitude for all sources with luminosities $> 10^4$\,\lsun\ (red histogram) and for the subset of these located between galactocentric radii of 5.5 and 6.5\,kpc (yellow histogram); this annulus effectively isolates sources associated with the Sagittarius arm. Inspection of the distribution of the whole sample reveals no evidence of an enhancement in the source density at the Sagittarius tangent. However, looking at the distribution of the subset of sources associated with the Sagittarius radius, there is an obvious increase in the source density towards the expected tangent position. The small numbers of sources in each longitude bin and, consequently, the relatively large errors make it hard to arrive at a definitive conclusion. However, this analysis shows that background contamination is a viable explanation.

\subsection{Galactic latitude distribution}

\begin{figure}
\includegraphics[width=0.90\linewidth]{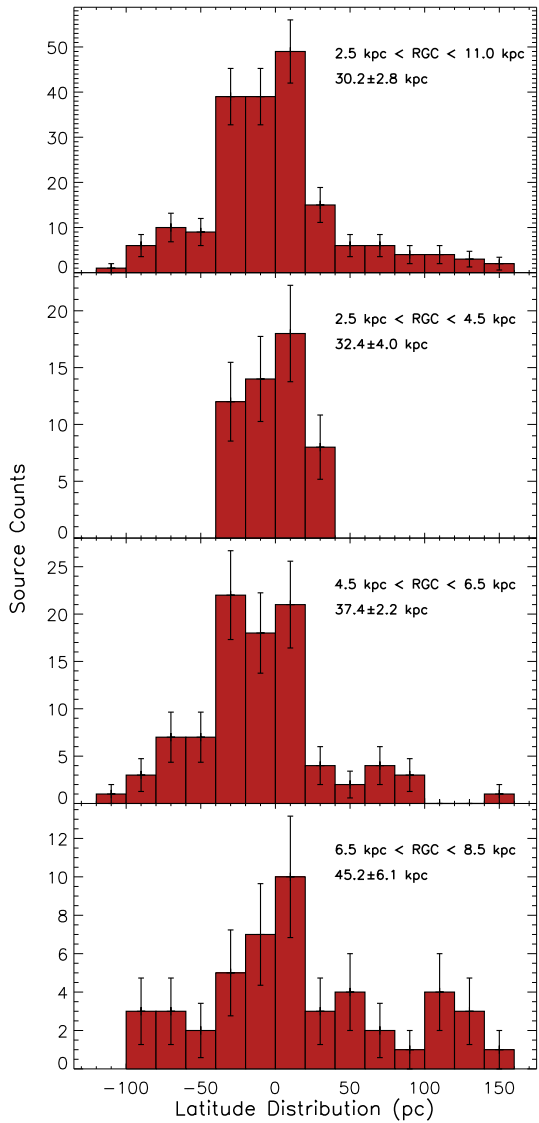}

\caption{\label{fig:gal_lat_dist}  The Galactocentric latitude distributions of the RMS sources plotted in the left panel. Errors have been calculated assuming \poi\ statistics.
}

\end{figure}

In Fig.\,\ref{fig:gal_lat_dist} we plot the distribution of young massive stars as a function of distance 
from the Galactic mid-plane. In the upper panel we present a histogram showing distribution of all RMS sources within the solar circle ($2.5\,{\rm{kpc}} < {\rm{RGC}} < 8.5\,{\rm{kpc}}$). The distribution of the whole sample peaks between 0--20\,pc above the plane, which is consistent with the value found by other authors (\citealt{reed2006}; and references therein). The latitude distribution is not particularly symmetrical, falling off more quickly above the plane than below. Thus, the scale height has slightly different values above and below the plane the scale heights are $\sim$17 and 42\,pc respectively. The skew is probably the result of our viewing angle from a distance above the plane of $\sim$\,20\,pc (\citealt{reed2006}).  Fitting a scale height to the whole data set we obtain an average value of 29$\pm0.5$\,pc, which agrees extremely well with the value measured by \citet{reed2000} from the distribution of nearby main sequence OB stars. 

In the lower three panels of Fig.\,\ref{fig:gal_lat_dist} we present histograms of the latitude distribution as a function of galactocentric radius (R$_{\rm GC}$). These plots reveal the Galactic latitude distribution to be dependent on R$_{\rm GC}$, with the scale height increasing with distance from the Galactic centre. The scale height increases from $\sim$30\,pc between $2.5\,{\rm{kpc}} < {\rm{R_{GC}}} < 4.5\,{\rm{kpc}}$ to $\sim$45\,pc between  $6.5\,{\rm{kpc}} < {\rm{R_{GC}}} < 8.5\,{\rm{kpc}}$ (see panels for scale heights obtained for the different galactocentric bins and their associated uncertainties). This increase in scale height with radius has also been seen in studies of the HI emission within the Milky Way (\citealt{malhotra1995}) and in external edge-on galaxies (\citealt{rupen1991}). The increase in the scale height as a function of galactocentric radius of both the young massive stars and HI is approximately linear. However, the scale height of the HI disk increases twice as quickly (from $\sim$100\,pc and $\sim$220\,pc between 3 and 8\,kpc).

In addition to the correlation between stellar scale height and disc radius, these plots also reveal that the centroid of the scale height oscillates above and below the Galactic midplane. This is again broadly in line with the distributions of both CO and HI emission (\citealt{malhotra1995} and references therein).

\section{Summary}
\label{sect:summary}

We have derived distances and luminosities to a large sample of MYSO candidates and UCHII regions identified by the Red MSX Source (RMS) survey. Distances to the majority of the sample have been obtained by cross-correlating with a sample of molecular clouds (i.e., $|b|<$1\degr\ and 17.8\degr\ $ < l <$  54.4\degr; \citealt{rathborne2009}) for which the distance ambiguities have been resolved using archival HI data (\citealt{roman2009}). In this way, we have obtained kinematic distances to 272 sources. Using these, we have calculated the Galactic scale height for  MYSOs and applied a 4$\sigma$ cut in order to break the distance ambiguity to high latitude sources located within the same longitude range (i.e., $|b|>$1\degr\ and 17.8\degr\ $ < l <$  54.4\degr). These two steps have provided kinematic distances to 292 sources out of a possible 326 located within this longitude range which corresponds to $\sim$90 per~cent of our sample. 

We used these assigned distances in conjunction with SED fits to calculate RMS source luminosities. We find 290 RMS sources have luminosities large enough to indicate the presence of a ZAMS star with a spectral type of B3 or earlier. From analysis of the luminosity distribution as a function of heliocentric distance, we estimate that our sample of young massive stars is complete for luminosities greater than $\sim$10$^4$\,\lsun\ out to a distance of $\sim$15\,kpc. In total we have identified 193 RMS sources with luminosities consistent with the presence of young massive stars above this completeness limit. 

We estimate the masses of 734 of the clouds in the GRS catalogue and, examining the mass-heliocentric distance distribution, we estimate that the catalogue is complete to clouds with masses greater than 10$^3$\,\msun\ within $\sim$15\,kpc. Selecting only clouds above this completeness limit and comparing the properties of those associated with young massive stars (i.e., $>10^4$\,\lsun) with the remainder of the clouds, we find significant differences. The clouds involved in massive star formation are, in general, larger, have higher column densities, have significantly larger masses and are more turbulent. Typical values for these parameters are  radius$\simeq$60\,pc, log(N(H$_2$))$\simeq$22 cm$^{-2}$, mass$\simeq$2.5$\times10^4$\,\msun\ and FWHM \vlsr$\simeq$4.3\,\kms.

The Galactocentric distribution of massive stars and molecular gas reveals strong peaks at approximately 4, 6 and 8\,kpc that correlate with streaming motions seen in HI data \citep{mcclure2007}, which is a strong indication that the mass and stellar density enhancements are associated with the spiral arms. Furthermore, the Galactic distribution of these young massive stars is found to be spatially correlated with the proposed locations of the Scutum, Sagittarius and Perseus spiral arms. We find a high concentration of stars with the location of the end of the Galactic bar and the Scutum spiral arm; this is also coincident with a bright region of molecular gas reported by \citet{roman2009}. There is a strong positional correlation between the massive stars and the Perseus and Sagittarius arms. We find similar levels of massive star formation associated with all three spiral arms. These results are consistent with a model of the Galaxy consisting of four principle arms. 

Using the latitude distribution of massive stars, we calculate the Galactic scale height for the whole sample of young massive stars to be $\sim29\pm0.5$\,pc, which is in excellent agreement with the value reported by \citet{reed2000} derived from a sample of nearby main sequence OB stars. However, measuring the scale height of the sample as a function of galactocentric radius we find the scale height increases with increasing distance from the Galactic centre.

In this paper we derive distances and luminosities to a sample of $\sim$300 MYSOs and UCHII regions identified from our programme of follow-up observations designed to examine the global characteristics of this galaxy-wide sample of massive young stars. Analysis of HI data is currently underway to derive distances and luminosities for a further $\sim$1000 sources; these results will be presented in a future publication. If the results found for the sample presented here are indicative of the remaining sources the RMS survey will fulfill its promise of returning the largest, well-selected, and Galaxy-wide sample of young massive stars yet produced.

\section*{Acknowledgments}

We are thankful to the anonymous referee for their thoughtful comments and suggestions which has significantly improved the quality of this work. JSU is supported by a CSIRO OCE postdoctoral grant. This publication makes use of molecular line data from the Boston University/FCRAO Galactic Ring Survey (GRS). The GRS is a joint project of Boston University and Five College Radio Astronomy Observatory, funded by the National Science Foundation under grants AST-9800334, AST-0098562, AST-0100793, AST-0228993, \& AST-0507657. This paper made use of information from the Red MSX Source survey database at www.ast.leeds.ac.uk/RMS which was constructed with support from the Science and Technology Facilities Council of the UK. This research makes use of data products from the MSX and 2MASS and GLIMPSE Surveys, which is are joint projects of the University of Massachusetts and the Infrared Processing and Analysis Center/California Institute of Technology, funded by the National Aeronautics and Space Administration and the National Science Foundation.

\bibliography{scuba2}

\begin{thebibliography}{}

\bibitem[\protect\citeauthoryear{{Alvarez}, {May} \& {Bronfman}}{{Alvarez}
  et~al.}{1990}]{alvarez1990}
{Alvarez} H.,  {May} J.,    {Bronfman} L.,  1990, \apj, 348, 495

\bibitem[\protect\citeauthoryear{{Anderson} \& {Bania}}{{Anderson} \&
  {Bania}}{2009}]{anderson2009a}
{Anderson} L.~D.,  {Bania} T.~M.,  2009, \apj, 690, 706

\bibitem[\protect\citeauthoryear{{Araya}, {Hofner}, {Churchwell} \&
  {Kurtz}}{{Araya} et~al.}{2001}]{araya2001}
{Araya} E.,  {Hofner} P.,  {Churchwell} E.,    {Kurtz} S.,  2001, in Bulletin
  of the American Astronomical Society Vol.~33, {Arecibo Observations of
  Formaldehyde and Radio Recombination Lines Toward Ultracompact HII Regions}.
p.~1487

\bibitem[\protect\citeauthoryear{{Araya}, {Hofner}, {Churchwell} \&
  {Kurtz}}{{Araya} et~al.}{2002}]{araya2002}
{Araya} E.,  {Hofner} P.,  {Churchwell} E.,    {Kurtz} S.,  2002, \apjs, 138,
  63

\bibitem[\protect\citeauthoryear{{Beltr{\'a}n}, {Brand}, {Cesaroni}, {Fontani},
  {Pezzuto}, {Testi} \& {Molinari}}{{Beltr{\'a}n} et~al.}{2006}]{beltran2006}
{Beltr{\'a}n} M.~T.,  {Brand} J.,  {Cesaroni} R.,  {Fontani} F.,  {Pezzuto} S.,
   {Testi} L.,    {Molinari} S.,  2006, \aap, 447, 221

\bibitem[\protect\citeauthoryear{{Benjamin}}{{Benjamin}}{2008}]{benjamin2008}
{Benjamin} R.~A.,  2008, in {H.~Beuther, H.~Linz, \& T.~Henning} ed., Massive
  Star Formation: Observations Confront Theory Vol.~387 of Astronomical Society
  of the Pacific Conference Series, {The Spiral Structure of the Galaxy:
  Something Old, Something New...}.
pp 375--+

\bibitem[\protect\citeauthoryear{{Benjamin}, {Churchwell}, {Babler},
  {Indebetouw}, {Meade}, {Whitney}, {Watson}, {Wolfire}, {Wolff}, {Ignace},
  {Bania}, {Bracker} \& {Clemens}}{{Benjamin} et~al.}{2005}]{benjamin2005}
{Benjamin} R.~A.,  {Churchwell} E.,  {Babler} B.~L.,  {Indebetouw} R.,  {Meade}
  M.~R.,  {Whitney} B.~A.,  {Watson} C.,  {Wolfire} M.~G.,  {Wolff} M.~J.,
  {Ignace} R.,  {Bania} T.~M.,  {Bracker} S.,    {Clemens} D.~P.,  2005, \apjl,
  630, L149

\bibitem[\protect\citeauthoryear{{Benjamin} \& et al.}{{Benjamin} \&
  et~al.}{2003}]{benjamin2003}
{Benjamin} R.~A.,  et al. 2003, \pasp, 115, 953

\bibitem[\protect\citeauthoryear{{Beuther}, {Walsh}, {Schilke}, {Sridharan},
  {Menten} \& {Wyrowski}}{{Beuther} et~al.}{2002}]{beuther2002}
{Beuther} H.,  {Walsh} A.,  {Schilke} P.,  {Sridharan} T.~K.,  {Menten} K.~M.,
    {Wyrowski} F.,  2002, \aap, 390, 289

\bibitem[\protect\citeauthoryear{{Binney}, {Gerhard}, {Stark}, {Bally} \&
  {Uchida}}{{Binney} et~al.}{1991}]{binney1991}
{Binney} J.,  {Gerhard} O.~E.,  {Stark} A.~A.,  {Bally} J.,    {Uchida} K.~I.,
  1991, \mnras, 252, 210

\bibitem[\protect\citeauthoryear{{Blitz} \& {Spergel}}{{Blitz} \&
  {Spergel}}{1991}]{blitz1991}
{Blitz} L.,  {Spergel} D.~N.,  1991, \apj, 379, 631

\bibitem[\protect\citeauthoryear{{Blitz} \& {Williams}}{{Blitz} \&
  {Williams}}{1999}]{blitz1999}
{Blitz} L.,  {Williams} J.~P.,  1999, ArXiv Astrophysics e-prints

\bibitem[\protect\citeauthoryear{{Bourke}, {Garay}, {Lehtinen}, {Koehnenkamp},
  {Launhardt}, {Nyman}, {May}, {Robinson} \& {Hyland}}{{Bourke}
  et~al.}{1997}]{bourke1997}
{Bourke} T.~L.,  {Garay} G.,  {Lehtinen} K.~K.,  {Koehnenkamp} I.,  {Launhardt}
  R.,  {Nyman} L.-A.,  {May} J.,  {Robinson} G.,    {Hyland} A.~R.,  1997,
  \apj, 476, 781

\bibitem[\protect\citeauthoryear{{Brand} \& {Blitz}}{{Brand} \&
  {Blitz}}{1993}]{brand1993}
{Brand} J.,  {Blitz} L.,  1993, \aap, 275, 67

\bibitem[\protect\citeauthoryear{{Braz} \& {Epchtein}}{{Braz} \&
  {Epchtein}}{1983}]{braz1983}
{Braz} M.~A.,  {Epchtein} N.,  1983, \aaps, 54, 167

\bibitem[\protect\citeauthoryear{{Bronfman}, {Nyman} \& {May}}{{Bronfman}
  et~al.}{1996}]{bronfman1996}
{Bronfman} L.,  {Nyman} L.-A.,    {May} J.,  1996, \aaps, 115, 81

\bibitem[\protect\citeauthoryear{{Busfield}, {Purcell}, {Hoare}, {Lumsden},
  {Moore} \& {Oudmaijer}}{{Busfield} et~al.}{2006}]{busfield2006}
{Busfield} A.~L.,  {Purcell} C.~R.,  {Hoare} M.~G.,  {Lumsden} S.~L.,  {Moore}
  T.~J.~T.,    {Oudmaijer} R.~D.,  2006, \mnras, 366, 1096

\bibitem[\protect\citeauthoryear{{Cao}, {Terebey}, {Prince} \&
  {Beichman}}{{Cao} et~al.}{1997}]{cao1997}
{Cao} Y.,  {Terebey} S.,  {Prince} T.~A.,    {Beichman} C.~A.,  1997, \apjs,
  111, 387

\bibitem[\protect\citeauthoryear{{Churchwell}, {Babler}, {Meade}, {Whitney},
  {Benjamin}, {Indebetouw}, {Cyganowski}, {Robitaille}, {Povich}, {Watson} \&
  {Bracker}}{{Churchwell} et~al.}{2009}]{churchwell2009}
{Churchwell} E.,  {Babler} B.~L.,  {Meade} M.~R.,  {Whitney} B.~A.,  {Benjamin}
  R.,  {Indebetouw} R.,  {Cyganowski} C.,  {Robitaille} T.~P.,  {Povich} M.,
  {Watson} C.,    {Bracker} S.,  2009, \pasp, 121, 213

\bibitem[\protect\citeauthoryear{{Clarke}, {Lumsden}, {Oudmaijer}, {Busfield},
  {Hoare}, {Moore}, {Sheret} \& {Urquhart}}{{Clarke} et~al.}{2006}]{clarke2006}
{Clarke} A.~J.,  {Lumsden} S.~L.,  {Oudmaijer} R.~D.,  {Busfield} A.~L.,
  {Hoare} M.~G.,  {Moore} T.~J.~T.,  {Sheret} T.~L.,    {Urquhart} J.~S.,
  2006, \aap, 457, 183

\bibitem[\protect\citeauthoryear{{Clemens}}{{Clemens}}{1985}]{clemens1985}
{Clemens} D.~P.,  1985, \apj, 295, 422

\bibitem[\protect\citeauthoryear{{Clemens}, {Sanders} \& {Scoville}}{{Clemens}
  et~al.}{1988}]{clemens1988}
{Clemens} D.~P.,  {Sanders} D.~B.,    {Scoville} N.~Z.,  1988, \apj, 327, 139

\bibitem[\protect\citeauthoryear{{Cohen}, {Hammersley} \& {Egan}}{{Cohen}
  et~al.}{2000}]{cohen2000}
{Cohen} M.,  {Hammersley} P.~L.,    {Egan} M.~P.,  2000, \aj, 120, 3362

\bibitem[\protect\citeauthoryear{{Cutri}, {Skrutskie}, {van Dyk}, {Beichman},
  {Carpenter}, {Chester}, {Cambresy}, {Evans}, {Fowler}, {Gizis}, {Howard},
  {Huchra}, {Jarrett}, {Kopan}, {Kirkpatrick}, {Light}, {Marsh} \&
  {McCallon}}{{Cutri} et~al.}{2003}]{cutri2003}
{Cutri} R.~M.,  {Skrutskie} M.~F.,  {van Dyk} S.,  {Beichman} C.~A.,
  {Carpenter} J.~M.,  {Chester} T.,  {Cambresy} L.,  {Evans} T.,  {Fowler} J.,
  {Gizis} J.,  {Howard} E.,  {Huchra} J.,  {Jarrett} T.,  {Kopan} E.~L.,
  {Kirkpatrick} J.~D.,  {Light} R.~M.,  {Marsh} K.~A.,    {McCallon} H.,  2003,
  VizieR Online Data Catalog, 2246, 0

\bibitem[\protect\citeauthoryear{{Dame}}{{Dame}}{1991}]{dame1991}
{Dame} T.~M.,  1991, in {A.~D.~Haschick \& P.~T.~P.~Ho} ed., Atoms, Ions and
  Molecules: New Results in Spectral Line Astrophysics Vol.~16 of Astronomical
  Society of the Pacific Conference Series, {Molecular Clouds as Tracers of
  Galactic Spiral Structure}.
pp 43--+

\bibitem[\protect\citeauthoryear{{Dame}}{{Dame}}{1993}]{dame1993}
{Dame} T.~M.,  1993, in {S.~S.~Holt \& F.~Verter} ed., Back to the Galaxy
  Vol.~278 of American Institute of Physics Conference Series, {The
  Distribution of Neutral Gas in the Milky Way}.
pp 267--278

\bibitem[\protect\citeauthoryear{{Dame}, {Hartmann} \& {Thaddeus}}{{Dame}
  et~al.}{2001}]{dame2001}
{Dame} T.~M.,  {Hartmann} D.,    {Thaddeus} P.,  2001, \apj, 547, 792

\bibitem[\protect\citeauthoryear{{De Buizer}, {Watson}, {Radomski}, {Pi{\~n}a}
  \& {Telesco}}{{De Buizer} et~al.}{2002}]{de-buizer2002}
{De Buizer} J.~M.,  {Watson} A.~M.,  {Radomski} J.~T.,  {Pi{\~n}a} R.~K.,
  {Telesco} C.~M.,  2002, \apjl, 564, L101

\bibitem[\protect\citeauthoryear{{de Wit}, {Testi}, {Palla}, {Vanzi} \&
  {Zinnecker}}{{de Wit} et~al.}{2004}]{de-wit2004}
{de Wit} W.~J.,  {Testi} L.,  {Palla} F.,  {Vanzi} L.,    {Zinnecker} H.,
  2004, \aap, 425, 937

\bibitem[\protect\citeauthoryear{{Di Francesco}, {Johnstone}, {Kirk},
  {MacKenzie} \& {Ledwosinska}}{{Di Francesco} et~al.}{2008}]{di-francesco2008}
{Di Francesco} J.,  {Johnstone} D.,  {Kirk} H.,  {MacKenzie} T.,
  {Ledwosinska} E.,  2008, \apjs, 175, 277

\bibitem[\protect\citeauthoryear{{Downes}, {Wilson}, {Bieging} \&
  {Wink}}{{Downes} et~al.}{1980}]{downes1980}
{Downes} D.,  {Wilson} T.~L.,  {Bieging} J.,    {Wink} J.,  1980, \aaps, 40,
  379

\bibitem[\protect\citeauthoryear{{Dwek}, {Arendt}, {Hauser}, {Kelsall},
  {Lisse}, {Moseley}, {Silverberg}, {Sodroski} \& {Weiland}}{{Dwek}
  et~al.}{1995}]{dwek1995}
{Dwek} E.,  {Arendt} R.~G.,  {Hauser} M.~G.,  {Kelsall} T.,  {Lisse} C.~M.,
  {Moseley} S.~H.,  {Silverberg} R.~F.,  {Sodroski} T.~J.,    {Weiland} J.~L.,
  1995, \apj, 445, 716

\bibitem[\protect\citeauthoryear{{Egan}, {Price}, {Kraemer}, {Mizuno}, {Carey},
  {Wright}, {Engelke}, {Cohen} \& {Gugliotti}}{{Egan} et~al.}{2003}]{egan2003}
{Egan} M.~P.,  {Price} S.~D.,  {Kraemer} K.~E.,  {Mizuno} D.~R.,  {Carey}
  S.~J.,  {Wright} C.~O.,  {Engelke} C.~W.,  {Cohen} M.,    {Gugliotti} M.~G.,
  2003, VizieR Online Data Catalog, 5114, 0

\bibitem[\protect\citeauthoryear{{Fa{\'u}ndez}, {Bronfman}, {Garay}, {Chini},
  {Nyman} \& {May}}{{Fa{\'u}ndez} et~al.}{2004}]{faundez2004}
{Fa{\'u}ndez} S.,  {Bronfman} L.,  {Garay} G.,  {Chini} R.,  {Nyman} L.,
  {May} J.,  2004, \aap, 426, 97

\bibitem[\protect\citeauthoryear{{Frerking}, {Langer} \& {Wilson}}{{Frerking}
  et~al.}{1982}]{frerking1982}
{Frerking} M.~A.,  {Langer} W.~D.,    {Wilson} R.~W.,  1982, \apj, 262, 590

\bibitem[\protect\citeauthoryear{{Georgelin} \& {Georgelin}}{{Georgelin} \&
  {Georgelin}}{1976}]{georgelin1976}
{Georgelin} Y.~M.,  {Georgelin} Y.~P.,  1976, \aap, 49, 57

\bibitem[\protect\citeauthoryear{{Hammersley}, {Garz{\'o}n}, {Mahoney},
  {L{\'o}pez-Corredoira} \& {Torres}}{{Hammersley}
  et~al.}{2000}]{hammersley2000}
{Hammersley} P.~L.,  {Garz{\'o}n} F.,  {Mahoney} T.~J.,  {L{\'o}pez-Corredoira}
  M.,    {Torres} M.~A.~P.,  2000, \mnras, 317, L45

\bibitem[\protect\citeauthoryear{{Hill}, {Burton}, {Minier}, {Thompson},
  {Walsh}, {Hunt-Cunningham} \& {Garay}}{{Hill} et~al.}{2005}]{hill2005}
{Hill} T.,  {Burton} M.~G.,  {Minier} V.,  {Thompson} M.~A.,  {Walsh} A.~J.,
  {Hunt-Cunningham} M.,    {Garay} G.,  2005, \mnras, 363, 405

\bibitem[\protect\citeauthoryear{{Hoare}, {Lumsden}, {Oudmaijer}, {Urquhart},
  {Busfield}, {Sheret}, {Clarke}, {Moore}, {Allsopp}, {Burton}, {Purcell},
  {Jiang} \& {Wang}}{{Hoare} et~al.}{2005}]{hoare2005}
{Hoare} M.~G.,  {Lumsden} S.~L.,  {Oudmaijer} R.~D.,  {Urquhart} J.~S.,
  {Busfield} A.~L.,  {Sheret} T.~L.,  {Clarke} A.~J.,  {Moore} T.~J.~T.,
  {Allsopp} J.,  {Burton} M.~G.,  {Purcell} C.~R.,  {Jiang} Z.,    {Wang} M.,
  2005, in {Cesaroni} R.,  {Felli} M.,  {Churchwell} E.,   {Walmsley} M.,  eds,
  Massive Star Birth: A Crossroads of Astrophysics Vol.~227 of IAU Symposium,
  {The RMS survey: Massive young stars throughout the galaxy}.
pp 370--375

\bibitem[\protect\citeauthoryear{{Hunter}, {Bertsch}, {Catelli}, {Dame},
  {Digel}, {Dingus}, {Esposito}, {Fichtel}, {Hartman}, {Kanbach}, {Kniffen},
  {Lin}, {Mayer-Hasselwander}, {Michelson}, {von Montigny}, {Mukherjee}
  et~al.,}{{Hunter} et~al.}{1997}]{hunter1997}
{Hunter} S.~D.,  {Bertsch} D.~L.,  {Catelli} J.~R.,  {Dame} T.~M.,  {Digel}
  S.~W.,  {Dingus} B.~L.,  {Esposito} J.~A.,  {Fichtel} C.~E.,  {Hartman}
  R.~C.,  {Kanbach} G.,  {Kniffen} D.~A.,  {Lin} Y.~C.,  {Mayer-Hasselwander}
  H.~A.,  {Michelson} P.~F.,  {von Montigny} C.,  {Mukherjee} R.,    et~al.,
  1997, \apj, 481, 205

\bibitem[\protect\citeauthoryear{{Jackson}, {Bania}, {Simon}, {Kolpak},
  {Clemens} \& {Heyer}}{{Jackson} et~al.}{2002}]{jackson2002}
{Jackson} J.~M.,  {Bania} T.~M.,  {Simon} R.,  {Kolpak} M.,  {Clemens} D.~P.,
   {Heyer} M.,  2002, \apjl, 566, L81

\bibitem[\protect\citeauthoryear{{Jackson}, {Finn}, {Rathborne}, {Chambers} \&
  {Simon}}{{Jackson} et~al.}{2008}]{jackson2008}
{Jackson} J.~M.,  {Finn} S.~C.,  {Rathborne} J.~M.,  {Chambers} E.~T.,
  {Simon} R.,  2008, \apj, 680, 349

\bibitem[\protect\citeauthoryear{{Kennicutt}}{{Kennicutt}}{2005}]{kennicutt200%
5}
{Kennicutt} R.~C.,  2005, in {Cesaroni} R.,  {Felli} M.,  {Churchwell} E.,
  {Walmsley} M.,  eds, Massive Star Birth: A Crossroads of Astrophysics
  Vol.~227 of IAU Symposium, {The role of massive stars in astrophysics}.
pp 3--11

\bibitem[\protect\citeauthoryear{{Kolpak}, {Jackson}, {Bania}, {Clemens} \&
  {Dickey}}{{Kolpak} et~al.}{2003}]{kolpak2003}
{Kolpak} M.~A.,  {Jackson} J.~M.,  {Bania} T.~M.,  {Clemens} D.~P.,    {Dickey}
  J.~M.,  2003, \apj, 582, 756

\bibitem[\protect\citeauthoryear{{Kuchar} \& {Bania}}{{Kuchar} \&
  {Bania}}{1994}]{kuchar1994}
{Kuchar} T.~A.,  {Bania} T.~M.,  1994, \apj, 436, 117

\bibitem[\protect\citeauthoryear{{Lekht}}{{Lekht}}{2000}]{lekht2000}
{Lekht} E.~E.,  2000, \aaps, 141, 185

\bibitem[\protect\citeauthoryear{{Lumsden}, {Hoare}, {Oudmaijer} \&
  {Richards}}{{Lumsden} et~al.}{2002}]{lumsden2002}
{Lumsden} S.~L.,  {Hoare} M.~G.,  {Oudmaijer} R.~D.,    {Richards} D.,  2002,
  \mnras, 336, 621

\bibitem[\protect\citeauthoryear{{Malhotra}}{{Malhotra}}{1995}]{malhotra1995}
{Malhotra} S.,  1995, \apj, 448, 138

\bibitem[\protect\citeauthoryear{{Martins}, {Schaerer} \& {Hillier}}{{Martins}
  et~al.}{2005}]{martins2005}
{Martins} F.,  {Schaerer} D.,    {Hillier} D.~J.,  2005, \aap, 436, 1049

\bibitem[\protect\citeauthoryear{{McClure-Griffiths} \&
  {Dickey}}{{McClure-Griffiths} \& {Dickey}}{2007}]{mcclure2007}
{McClure-Griffiths} N.~M.,  {Dickey} J.~M.,  2007, \apj, 671, 427

\bibitem[\protect\citeauthoryear{{Meynet} \& {Maeder}}{{Meynet} \&
  {Maeder}}{2000}]{meynet2000}
{Meynet} G.,  {Maeder} A.,  2000, \aap, 361, 101

\bibitem[\protect\citeauthoryear{{Mottram}, {Hoare}, {Lumsden}, {Oudmaijer},
  {Urquhart}, {Meade}, {Moore} \& {Stead}}{{Mottram}
  et~al.}{2010}]{mottram2009}
{Mottram} J.~C.,  {Hoare} M.~G.,  {Lumsden} S.~L.,  {Oudmaijer} R.~D.,
  {Urquhart} J.~S.,  {Meade} M.~R.,  {Moore} T.~J.~T.,    {Stead} J.~J.,  2010,
  \aap, 510, 89

\bibitem[\protect\citeauthoryear{{Mottram}, {Hoare}, {Lumsden}, {Oudmaijer},
  {Urquhart}, {Sheret}, {Clarke} \& {Allsopp}}{{Mottram}
  et~al.}{2007}]{mottram2007}
{Mottram} J.~C.,  {Hoare} M.~G.,  {Lumsden} S.~L.,  {Oudmaijer} R.~D.,
  {Urquhart} J.~S.,  {Sheret} T.~L.,  {Clarke} A.~J.,    {Allsopp} J.,  2007,
  \aap, 476, 1019

\bibitem[\protect\citeauthoryear{{Paladini}, {Davies} \& {De Zotti}}{{Paladini}
  et~al.}{2004}]{paladini2004}
{Paladini} R.,  {Davies} R.~D.,    {De Zotti} G.,  2004, \mnras, 347, 237

\bibitem[\protect\citeauthoryear{{Price}, {Egan}, {Carey}, {Mizuno} \&
  {Kuchar}}{{Price} et~al.}{2001}]{price2001}
{Price} S.~D.,  {Egan} M.~P.,  {Carey} S.~J.,  {Mizuno} D.~R.,    {Kuchar}
  T.~A.,  2001, \aj, 121, 2819

\bibitem[\protect\citeauthoryear{{Rathborne}, {Johnson}, {Jackson}, {Shah} \&
  {Simon}}{{Rathborne} et~al.}{2009}]{rathborne2009}
{Rathborne} J.~M.,  {Johnson} A.~M.,  {Jackson} J.~M.,  {Shah} R.~Y.,
  {Simon} R.,  2009, \apjs, 182, 131

\bibitem[\protect\citeauthoryear{{Reed}}{{Reed}}{2000}]{reed2000}
{Reed} B.~C.,  2000, \aj, 120, 314

\bibitem[\protect\citeauthoryear{{Reed}}{{Reed}}{2006}]{reed2006}
{Reed} B.~C.,  2006, \jrasc, 100, 146

\bibitem[\protect\citeauthoryear{{Robitaille}, {Whitney}, {Indebetouw} \&
  {Wood}}{{Robitaille} et~al.}{2007}]{robitaille2007}
{Robitaille} T.~P.,  {Whitney} B.~A.,  {Indebetouw} R.,    {Wood} K.,  2007,
  \apjs, 169, 328

\bibitem[\protect\citeauthoryear{{Robitaille}, {Whitney}, {Indebetouw}, {Wood}
  \& {Denzmore}}{{Robitaille} et~al.}{2006}]{robitaille2006}
{Robitaille} T.~P.,  {Whitney} B.~A.,  {Indebetouw} R.,  {Wood} K.,
  {Denzmore} P.,  2006, \apjs, 167, 256

\bibitem[\protect\citeauthoryear{{Roman-Duval}, {Jackson}, {Heyer}, {Johnson},
  {Rathborne}, {Shah} \& {Simon}}{{Roman-Duval} et~al.}{2009}]{roman2009}
{Roman-Duval} J.,  {Jackson} J.~M.,  {Heyer} M.,  {Johnson} A.,  {Rathborne}
  J.,  {Shah} R.,    {Simon} R.,  2009, \apj, 699, 1153

\bibitem[\protect\citeauthoryear{{Rupen}}{{Rupen}}{1991}]{rupen1991}
{Rupen} M.~P.,  1991, \aj, 102, 48

\bibitem[\protect\citeauthoryear{{Sanders}, {Clemens}, {Scoville} \&
  {Solomon}}{{Sanders} et~al.}{1986}]{sanders1986}
{Sanders} D.~B.,  {Clemens} D.~P.,  {Scoville} N.~Z.,    {Solomon} P.~M.,
  1986, \apjs, 60, 1

\bibitem[\protect\citeauthoryear{{Sewilo}, {Watson}, {Araya}, {Churchwell},
  {Hofner} \& {Kurtz}}{{Sewilo} et~al.}{2004}]{sewilo2004}
{Sewilo} M.,  {Watson} C.,  {Araya} E.,  {Churchwell} E.,  {Hofner} P.,
  {Kurtz} S.,  2004, \apjs, 154, 553

\bibitem[\protect\citeauthoryear{{Simon}, {Jackson}, {Bania}, {Clemens},
  {Heyer}, {Egan} \& {Price}}{{Simon} et~al.}{2001}]{simon2001}
{Simon} R.,  {Jackson} J.~M.,  {Bania} T.~M.,  {Clemens} D.~P.,  {Heyer} M.~H.,
   {Egan} M.~P.,    {Price} S.~D.,  2001, in Bulletin of the American
  Astronomical Society Vol.~33 of Bulletin of the American Astronomical
  Society, {Infrared Dark Clouds in the BU-FCRAO Milky Way Galactic Ring
  Survey}.
pp 1450--+

\bibitem[\protect\citeauthoryear{{Skrutskie} \& et al.}{{Skrutskie} \&
  et~al.}{2006}]{skrutskie2006}
{Skrutskie} M.~F.,  et al. 2006, \aj, 131, 1163

\bibitem[\protect\citeauthoryear{{Solomon}, {Rivolo}, {Barrett} \&
  {Yahil}}{{Solomon} et~al.}{1987}]{solomon1987}
{Solomon} P.~M.,  {Rivolo} A.~R.,  {Barrett} J.,    {Yahil} A.,  1987, \apj,
  319, 730

\bibitem[\protect\citeauthoryear{{Stil}, {Taylor}, {Dickey}, {Kavars},
  {Martin}, {Rothwell}, {Boothroyd}, {Lockman} \& {McClure-Griffiths}}{{Stil}
  et~al.}{2006}]{stil2006}
{Stil} J.~M.,  {Taylor} A.~R.,  {Dickey} J.~M.,  {Kavars} D.~W.,  {Martin}
  P.~G.,  {Rothwell} T.~A.,  {Boothroyd} A.~I.,  {Lockman} F.~J.,
  {McClure-Griffiths} N.~M.,  2006, \aj, 132, 1158

\bibitem[\protect\citeauthoryear{{Urquhart}, {Busfield}, {Hoare}, {Lumsden},
  {Clarke}, {Moore}, {Mottram} \& {Oudmaijer}}{{Urquhart}
  et~al.}{2007}]{urquhart_radio_south}
{Urquhart} J.~S.,  {Busfield} A.~L.,  {Hoare} M.~G.,  {Lumsden} S.~L.,
  {Clarke} A.~J.,  {Moore} T.~J.~T.,  {Mottram} J.~C.,    {Oudmaijer} R.~D.,
  2007, \aap, 461, 11

\bibitem[\protect\citeauthoryear{{Urquhart}, {Busfield}, {Hoare}, {Lumsden},
  {Oudmaijer}, {Moore}, {Gibb}, {Purcell}, {Burton} \& {Marechal}}{{Urquhart}
  et~al.}{2007}]{urquhart_13co_south}
{Urquhart} J.~S.,  {Busfield} A.~L.,  {Hoare} M.~G.,  {Lumsden} S.~L.,
  {Oudmaijer} R.~D.,  {Moore} T.~J.~T.,  {Gibb} A.~G.,  {Purcell} C.~R.,
  {Burton} M.~G.,    {Marechal} L.~J.~L.,  2007, \aap, 474, 891

\bibitem[\protect\citeauthoryear{{Urquhart}, {Busfield}, {Hoare}, {Lumsden},
  {Oudmaijer}, {Moore}, {Gibb}, {Purcell}, {Burton}, {Mar{\'e}chal}, {Jiang} \&
  {Wang}}{{Urquhart} et~al.}{2008}]{urquhart_13co_north}
{Urquhart} J.~S.,  {Busfield} A.~L.,  {Hoare} M.~G.,  {Lumsden} S.~L.,
  {Oudmaijer} R.~D.,  {Moore} T.~J.~T.,  {Gibb} A.~G.,  {Purcell} C.~R.,
  {Burton} M.~G.,  {Mar{\'e}chal} L.~J.~L.,  {Jiang} Z.,    {Wang} M.,  2008,
  \aap, 487, 253

\bibitem[\protect\citeauthoryear{{Urquhart}, {Hoare}, {Lumsden}, {Oudmaijer} \&
  {Moore}}{{Urquhart} et~al.}{2008}]{urquhart2007c}
{Urquhart} J.~S.,  {Hoare} M.~G.,  {Lumsden} S.~L.,  {Oudmaijer} R.~D.,
  {Moore} T.~J.~T.,  2008, in {Beuther} H.,  {Linz} H.,   {Henning} T.,  eds,
  Massive Star Formation: Observations Confront Theory Vol.~387 of Astronomical
  Society of the Pacific Conference Series, {The RMS Survey: A Galaxy-wide
  Sample of Massive Young Stellar Objects}.
pp 381--+

\bibitem[\protect\citeauthoryear{{Urquhart}, {Hoare}, {Lumsden}, {Oudmaijer},
  {Moore}, {Brook}, {Mottram}, {Davies} \& {Stead}}{{Urquhart}
  et~al.}{2009}]{urquhart2009_h2o}
{Urquhart} J.~S.,  {Hoare} M.~G.,  {Lumsden} S.~L.,  {Oudmaijer} R.~D.,
  {Moore} T.~J.~T.,  {Brook} P.~R.,  {Mottram} J.~C.,  {Davies} B.,    {Stead}
  J.~J.,  2009, \aap, 507, 795

\bibitem[\protect\citeauthoryear{{Urquhart}, {Hoare}, {Purcell}, {Lumsden},
  {Oudmaijer}, {Moore}, {Busfield}, {Mottram} \& {Davies}}{{Urquhart}
  et~al.}{2009}]{urquhart_radio_north}
{Urquhart} J.~S.,  {Hoare} M.~G.,  {Purcell} C.~R.,  {Lumsden} S.~L.,
  {Oudmaijer} R.~D.,  {Moore} T.~J.~T.,  {Busfield} A.~L.,  {Mottram} J.~C.,
  {Davies} B.,  2009, \aap, 501, 539

\bibitem[\protect\citeauthoryear{{Williams}, {de Geus} \& {Blitz}}{{Williams}
  et~al.}{1994}]{williams1994}
{Williams} J.~P.,  {de Geus} E.~J.,    {Blitz} L.,  1994, \apj, 428, 693

\bibitem[\protect\citeauthoryear{{Wood} \& {Churchwell}}{{Wood} \&
  {Churchwell}}{1989}]{wood1989}
{Wood} D.~O.~S.,  {Churchwell} E.,  1989, \apjs, 69, 831

\bibitem[\protect\citeauthoryear{{Xu}, {Reid}, {Zheng} \& {Menten}}{{Xu}
  et~al.}{2006}]{xu2006}
{Xu} Y.,  {Reid} M.~J.,  {Zheng} X.~W.,    {Menten} K.~M.,  2006, Science, 311,
  54

\end{thebibliography}

\bibliographystyle{mn2e}

\bsp


\end{document}